\documentclass[12pt,preprint]{aastex}

\newcommand{\etal}{et\,al.}
\newcommand{\halpha}{H$\alpha$}
\newcommand{\lsim}{\raise0.3ex\hbox{$<$}\kern-0.75em{\lower0.65ex\hbox{$\sim$}}}

\newcommand{\msun}{M$_{\odot}$}

\newcommand{\HI}{H~{\sc I}}
\newcommand{\kms}{km\,s$^{-1}$}
\newcommand{\pom}{\,$\pm$\,}

\newcommand{\gsim}{\raise0.3ex\hbox{$>$}\kern-0.75em{\lower0.65ex\hbox{$\sim$}}}
\begin{document}
\slugcomment{The Astrophysical Journal, in press}

\title{The M81 Group Dwarf Irregular Galaxy DDO 165. I.  High Velocity
Neutral Gas in a Post-Starburst System}




\author{John M. Cannon, Hans P. Most}
\affil{Department of Physics \& Astronomy, Macalester College, 1600 Grand Avenue, Saint Paul, MN 55105}
\email{jcannon@macalester.edu}

\author{Evan D. Skillman, Daniel R. Weisz}
\affil{Astronomy Department, University of Minnesota, Minneapolis, MN 55455}
\email{skillman@astro.umn.edu, dweisz@astro.umn.edu}

\author{David Cook}
\affil{Department of Physics and Astronomy, University of Wyoming, Laramie, WY 82071, USA}
\email{dcook12@uwyo.edu}

\author{Andrew E. Dolphin}
\affil{Raytheon Company, 1151 East Hermans Road, Tucson, AZ 85756}
\email{adolphin@raytheon.com}

\author{Robert C. Kennicutt, Jr.}
\affil{Institute of Astronomy, University of Cambridge, Madingley Road,
Cambridge, CB3 0HA, United Kingdom}
\email{robk@ast.cam.ac.uk}

\author{Janice Lee}
\affil{Observatories of the Carnegie Institution of Washington, 813 Santa Barbara Street, Pasadena, CA 91101}
\email{jlee@obs.carnegiescience.edu}

\author{Anil Seth\footnote{OIR Fellow}}
\affil{Harvard-Smithsonian Center for Astrophysics, 60 Garden Street Cambridge, MA 02138}
\email{aseth@cfa.harvard.edu}

\author{Fabian Walter}
\affil{Max-Planck-Institut f{\"u}r Astronomie, K{\"o}nigstuhl 17, D-69117, Heidelberg, Germany}
\email{walter@mpia.de}

\author{Steven R. Warren}
\affil{Astronomy Department, University of Minnesota, Minneapolis, MN 55455}
\email{warren@astro.umn.edu}

\begin{abstract}

We present new multi-configuration {\it VLA} \HI\ spectral line
observations of the M81 group drr post-starburst galaxy DDO\,165. The
\HI\ morphology is complex, with multiple column density peaks
surrounding a large region of very low \HI\ surface density that is
offset from the center of the stellar distribution. The bulk of the
neutral gas is associated with the southern section of the galaxy; a
secondary peak in the north contains $\sim$15\% of the total
\HI\ mass. These components appear to be kinematically distinct,
suggesting that either tidal processes or large-scale blowout have
recently shaped the ISM of DDO\,165. Using spatially-resolved
position-velocity maps, we find multiple localized high-velocity gas
features. Cross-correlating with radius-velocity analyses, we identify
eight shell/hole structures in the ISM with a range of sizes
($\sim$400--900 pc) and expansion velocities ($\sim$7--11 km
s$^{-1}$). These structures are compared with narrow- and broad-band
imaging from {\it KPNO} and {\it HST}. Using the latter data, recent
works have shown that DDO 165's previous ``burst'' phase was extended
temporally ($\gsim$1 Gyr). We thus interpret the high-velocity gas
features, \HI\ holes, and kinematically distinct components of the
galaxy in the context of the immediate effects of ``feedback'' from
recent star formation.  In addition to creating \HI\ holes and shells,
extended star formation events are capable of creating localized high
velocity motion of the surrounding interstellar material.  A companion
paper connects the energetics from the \HI\ and {\it HST} data.

\end{abstract}						


\section{Introduction}
\label{S1}

Dwarf irregular (dIrr) galaxies are commonly assumed to be ``simple''
systems.  Compared with more massive star-forming galaxies, such
low-mass dwarfs lack bulk sources of energy that can ``trigger'' star
formation (e.g., spiral density waves, rotational shear).  Rather,
dIrrs are often observed undergoing solid-body rotation (see
discussion in {Skillman 1996}\nocite{skillman96} and numerous recent
examples; e.g., {Begum \etal\ 2008}\nocite{begum08}).  This implies
that star formation (SF) proceeds largely in situ, with
turbulence and stellar evolution being important dynamical processes.

This deceiving simplicity raises an important unanswered question in
galaxy evolution: how do relatively isolated, low-mass systems sustain
SF for extended periods of time?  Recent resolved stellar population
work has shown that dwarf galaxies can host remarkably long episodes
of heightened SF activity.  The durations of these bursts of SF are
commonly defined by a modified birthrate parameter, {\it b} $=$
SFR/$<$SFR$_{\rm past}>$; \citet{kennicutt05} sets a threshold of {\it
b}$>$2 as a ``starburst''.  Using this metric, burst durations of
order 100 Myr, and extending to $>$1 Gyr in some cases, have been
derived for nearby dwarfs ({Weisz \etal\ 2008}\nocite{weisz08};
{McQuinn \etal\ 2009}\nocite{mcquinn09}, {2010a}\nocite{mcquinn10a},
{2010b}\nocite{mcquinn10b}).  Further, these works have shown that
relatively young stars ($\lsim$100 Myr) can occupy a large fraction of
the surface area of the galaxy: the SF events can, at times, pervade
the entire disk.  These points serve as pertinent reminders that the
commonly-used instantaneous SF tracers (e.g., \halpha\ emission) are
only a small piece of a much larger, temporally extended, puzzle
\citep[see, e.g., the detailed discussion in][]{lee10}.

The current and past SF traced by these spatially resolved stellar
population studies naturally leads to mechanical energy input into the
interstellar medium (ISM) in the form of supernovae (SNe) and stellar
winds (hereafter collectively referred to as ``feedback'').  While
dwarf galaxies have small potential wells (dynamical masses are
typically of order 10$^7$--10$^9$ \msun; e.g., {Mateo
1998}\nocite{mateo98}), simulations suggest that it is difficult to
eject material from low-mass galaxies completely. \citet{maclow99} and
\citet{ferrara00} find that only systems with gas masses $\lsim$10$^7$
\msun\ are susceptible to large-scale mass loss from mechanical
feedback.  Multi-phase outflows of gas that apparently remain bound to
the systems, however, are predicted theoretically (see references
above) and verified observationally (e.g., {Heckman \etal\
2001}\nocite{heckman01}; {Martin \etal\ 2002}\nocite{martin02};
{Cannon \etal\ 2004}\nocite{cannon04}; {Young \etal\
2007}\nocite{young07}; {Kobulnicky \& Skillman
2008}\nocite{kobulnicky08}).

A parallel question to that posed above is what mechanism(s), if any,
serve(s) to regulate the ongoing SF in dwarfs?  What properties are
found in systems that have hosted a previous burst but now are
comparatively quiescent?  While such ``post-starburst'' galaxies are
not as well-studied as systems hosting ongoing bursts (see, for
example, the sample presented in {Lee \etal\ 2009}\nocite{lee09}),
they offer a unique opprtunity to study the stellar and gaseous
components of systems after vigorous SF episodes have terminated.
What halts or slows the SF in these systems?  If they are sufficiently
massive to maintain most or all of their gaseous component during a SF
event, then the cessation of SF cannot be caused by complete
``blowaway'' of the galaxies' available fuel.

One decidedly simplistic potential solution is that mechanical
``feedback'' slowly but cumulatively increases the velocity dispersion
and turbulence of the neutral gas \citep[e.g.,][]{joung06}.  This can
heat and distribute the gas over a sufficiently large volume to slow
or halt SF.  After this energy dissipates in the ISM, the neutral gas
will eventually begin to move downward in the potential well of the
galaxy.  SF would then re-ignite in Jeans-unstable regions.  In this
simplistic scenario, one would expect localized regions of SF to
impact their surroundings as the massive stars contained within
undergo the stellar life cycle \citep[e.g.,][]{stinson09}.

Some dwarf galaxies show abundant evidence for this SF - ISM
interaction. The ISM of the M81 group dwarf IC\,2574 is littered with
holes, shells and depressions \citep{walter99}; some of these are
associated with remnant stellar clusters within (e.g., {Stewart \&
Walter 2000}\nocite{stewart00}; {Cannon
\etal\ 2005}\nocite{cannon05}), and some are not.  Further, some of
these kinematic features directly correlate with both high column
density neutral gas and with high surface brightness nebular emission.
Recent spatially resolved work has shown that SF is energetically and
temporally capable of creating the ``Super Giant Shell'' structure,
and of igniting subsequent SF around the shell rim
\citep{cannon05,weisz09a}.  Similarly, in the M81 group dIrr galaxy
Holmberg\,II, temporally extended SF was found to be energetically
capable of creating the numerous holes in the ISM ({Weisz
\etal\ 2009b}\nocite{weisz09b}; see also {Puche
\etal\ 1992}\nocite{puche92} and {Rhode
\etal\ 1999}\nocite{rhode99}).  Taken together, these results
suggest that concentrated SF (i.e., ``bursts'') and subsequent stellar
evolution can (but do not always) lead to the formation of prominent
structures in the ISM of dwarf galaxies.

The scenarios discussed above raise another question that is important
for understanding the evolution of dwarf galaxies: over what
timescales do we expect to find correlations between intense SF, high
column density neutral gas (i.e., the ``Schmidt'' law; {Kennicutt
1989}\nocite{kennicutt89}, {1998}\nocite{kennicutt98}), and the
``canonical'' SF tracers (e.g., \halpha\ emission)?  Numerous works
have shown that current SF correlates well with emission in
various wavebands \citep[e.g., UV, optical and infrared, or
combinations thereof; ][]{calzetti05,kennicutt07,bigiel10}.  However, we do not
yet know if similar correlations hold true for SF intensity 100 Myr
ago, 200 Myr ago, and so on.  The aforementioned stellar population
studies, combined with multiwavelength imaging, provide an opportunity
to address this important issue.

In the present work we discuss these science themes as applied to the
intriguing dwarf galaxy DDO\,165 (a.k.a. UGC 8201).  Located on the far
side of the M81 group at a distance of 4.47\,\pom\,0.20 Mpc \citep[using
the magnitude of the tip of the red giant branch;][]{karachentsev02},
this relatively isolated system (tidal index $\Theta$ $=$ 0.0;
{Karachentsev \etal\ 2004}\nocite{karachentsev04}) has drawn attention
due to its peculiar properties.  A sharp boundary or cutoff of stars
distinguishes the galaxy's optical appearance.  \citet{stil02a} and
the {\it WHISP} survey \citep{kamphuis96} noted that the \HI\
kinematics were highly disturbed.  A member of the {\it SINGS} sample
\citep{kennicutt03}, the metal abundance in the ISM of DDO\,165 is
sub-solar ($\sim$13\% Z$_{\odot}$; see {Walter \etal\
2007}\nocite{walter07}, {Croxall \etal\ 2009}\nocite{croxall09}).
This, in part, contributes to the very weak mid- and far-infrared
emission (see images in {Walter \etal\ 2007}\nocite{walter07});
\citet{draine07} do not estimate a dust mass based on the marginal
detections at 70 and 160 $\mu$m using {\it Spitzer} imaging.

The narrowband studies by {Lee \etal\ (2007}\nocite{lee07},
{2009)}\nocite{lee09}, and the stellar population work by
\citet{weisz08}, have identified DDO\,165 as being in a post-starburst
state.  The latter study provided SF intensities as a function of time
throughout the galaxy back to $\sim$1 Gyr, and a global average value
over the Hubble time.  This lifetime average star formation rate (SFR)
was found to be a relatively quiescent $\sim$0.013 \msun\,yr$^{-1}$;
however, over the last 1 Gyr, DDO\,165 has formed 17\% of its total
stellar population.  Further, during this last Gyr, the SFR was
significantly elevated above the lifetime average (that is, the galaxy
was ``bursting''; see {McQuinn \etal\ 2010b}\nocite{mcquinn10b}).
Around 25 Myr ago, however, the global SFR plummeted to the current
low level; weak \halpha\ emission (see further discussion below)
verifies this dramatic reduction in current massive SFR.

Given these intriguing properties, we wish to ascertain what effects
the recent ($\sim$500 Myr ago to present) SF has had on the ISM of
DDO\,165.  Further, we seek clues as to what mechanism(s) have
contributed to the decreasing recent SFR.  To this end, we present new
\HI\ spectral line imaging of this system, and compare these rich data
with optical ground- and space-based imaging.  We have discovered a
wealth of morphological and kinematic structure in the ISM of
DDO\,165; we discuss these features in detail in this work.  A
companion paper uses spatially resolved stellar population analyses to
quantitatively study the stellar-ISM interaction over the last 500 Myr
in DDO\,165; there we find that SF has occurred throughout the entire
disk over this interval.

\section{Observations and Data Reduction}
\label{S2}

\subsection{VLA Spectral Line Observations}
\label{S2.1}

DDO\,165 (see basic parameters in Table~\ref{t1}) was observed with
the NRAO Very Large Array\footnote{The National Radio Astronomy
  Observatory is a facility of the National Science Foundation
  operated under cooperative agreement by Associated Universities,
  Inc.}  (VLA) in the \HI\ spectral line for program AC842 (PI
Cannon).  To be sensitive to \HI\ emission on various scales,
observations were acquired in the B, C, and D configurations (see
discussion in {Walter \etal\ 2008}\nocite{walter08}).  The VLA was
being upgraded to the {\it EVLA} during this program; we thus used a
set frequency to acquire all data (i.e., Doppler tracking was not
applied).  The on-source integration times were approximately 6.5 hr
in the B array, 3 hr in the C array, and 1.5 hr in the D array.  A
bandwidth of 1.56 MHz was used over 128 channels, resulting in a
velocity resolution of 2.58 \kms\ after online Hanning smoothing.
Table~\ref{t2} shows the observational parameters for each of the
observation sessions.  The primary calibrator 1331$+$305 (3c286) was
used for flux calibration\footnote{See
  http://www.vla.nrao.edu/astro/calib/manual/index.shtml}; the
secondary calibrator was 1313$+$675, with a flux density of 2.40 Jy.

We reduced the data from each observation separately using the
Astronomical Image Processing System (AIPS\footnote{The Astronomical
Image Processing System (AIPS) has been developed by the NRAO.})
package. First, the data were inspected and bad data points removed.
Flux, gain, and phase calibrations were then applied.  All three data
sets were then combined and imaged.  Line emission was prominent
between $-$20 \kms\ and $+$80 \kms.  The task IMAGR in AIPS was used
to clean to 2.5 times the RMS noise level.  This task was conducted
twice, once to create a naturally weighted data cube with an output
beam size of 12.5\arcsec\ $\times$ 11.3\arcsec\ ($\sigma$ $=$ 0.59
mJy\,Bm$^{-1}$) and a second time to create a robust weighted cube
(ROBUST = 0.5; {Briggs 1995}\nocite{briggs95}) with an output beam
size of 7.5\arcsec\ $\times$ 5.9\arcsec\ ($\sigma$ $=$ 0.67
mJy\,Bm$^{-1}$).  We convolved these cubes to four different circular
beam sizes: 20\arcsec\ $\times$ 20\arcsec\ and 15\arcsec\ $\times$
15\arcsec\ from the naturally weighted cube, and 10\arcsec\ $\times$
10\arcsec\ and 7.5\arcsec\ $\times$ 7.5\arcsec\ from the robust
weighted cube.  We explicitly account for the different beam sizes of
the dirty and clean maps using residual flux scaling (e.g.,
{J\"ors\"ater \& van Moorsel 1995}\nocite{jorsater95}; {Walter
\etal\ 2007}\nocite{walter07}).

To differentiate noise from real emission, the naturally weighted
20\arcsec\ cube was spatially smoothed and then blanked at the
2$\sigma$ level.  After blanking, the cube was further inspected by
hand.  Emission was classified as real when it appeared in three or
more consecutive channels.  This master blanked cube was then used to
blank all cubes, ensuring that the same regions contribute at each
resolution.  Using the blanked cubes, \HI\ surface brightness,
velocity field, and velocity dispersion maps were created using the
task XMOM in AIPS.

\subsection{Optical Observations}
\label{S2.2}

\halpha\ observations of DDO\,165 were obtained with the Bok 2.3\,m
telescope at Kitt Peak National Observatory (KPNO) on May 29, 2001.
The 2K CCD imager was used for 1000 seconds of on-source exposure
time.  The CCD read noise was 8.5 e$^-$ and the gain was set to 3.4
e$^-$ ADU$^{-1}$. An 88\,mm Andover 3-cavity interference filter was
used; this filter is sensitive to both \halpha\ and to [NII] emission.
The \halpha\ and R-band data were originally presented in
\citet{kennicutt08}, and detailed image reduction and calibration
information can be found there.  The 3$\sigma$ detection threshold is
$\sim$2\,$\times$\,10$^{-18}$ erg\,s$^{-1}$\,cm$^{-2}$.

R-band observations of DDO\,165 were obtained with the KPNO 2.1\,m
telescope on April 13, 2002.  The T2KA detector was used for 420
seconds of on-source integration time. The CCD read noise was 4 e$^-$
and the gain was set to 3.6 e$^-$ ADU$^{-1}$.  The image was reduced
following standard procedures using IRAF\footnote{The Image Reduction
and Analysis Facility (IRAF) is distributed by the National Optical
Astronomy Observatories, which are operated by AURA, Inc. under
cooperative agreement with the National Science Foundation.}.  The 
data are also discussed in \citet{kennicutt08}.

{\it HST} observations of DDO\,165 were obtained using the Advanced
Camera for Surveys (ACS) on March 17, 2006.  F555W and F814W filters
were used.  The data were processed by the standard {\it HST}
pipeline.  The V and I band photometry was presented in
\citet{weisz08}; the integration times were 9536 seconds in each
filter.  The treatment of the {\it HST} data are discussed in detail
in Paper II.

\section{Multiwavelength Properties of DDO\,165}
\label{S3}

\subsection{\HI\ Morphology and Dynamics}
\label{S3.1}

\begin{figure}
\epsscale{1.0}
\plotone{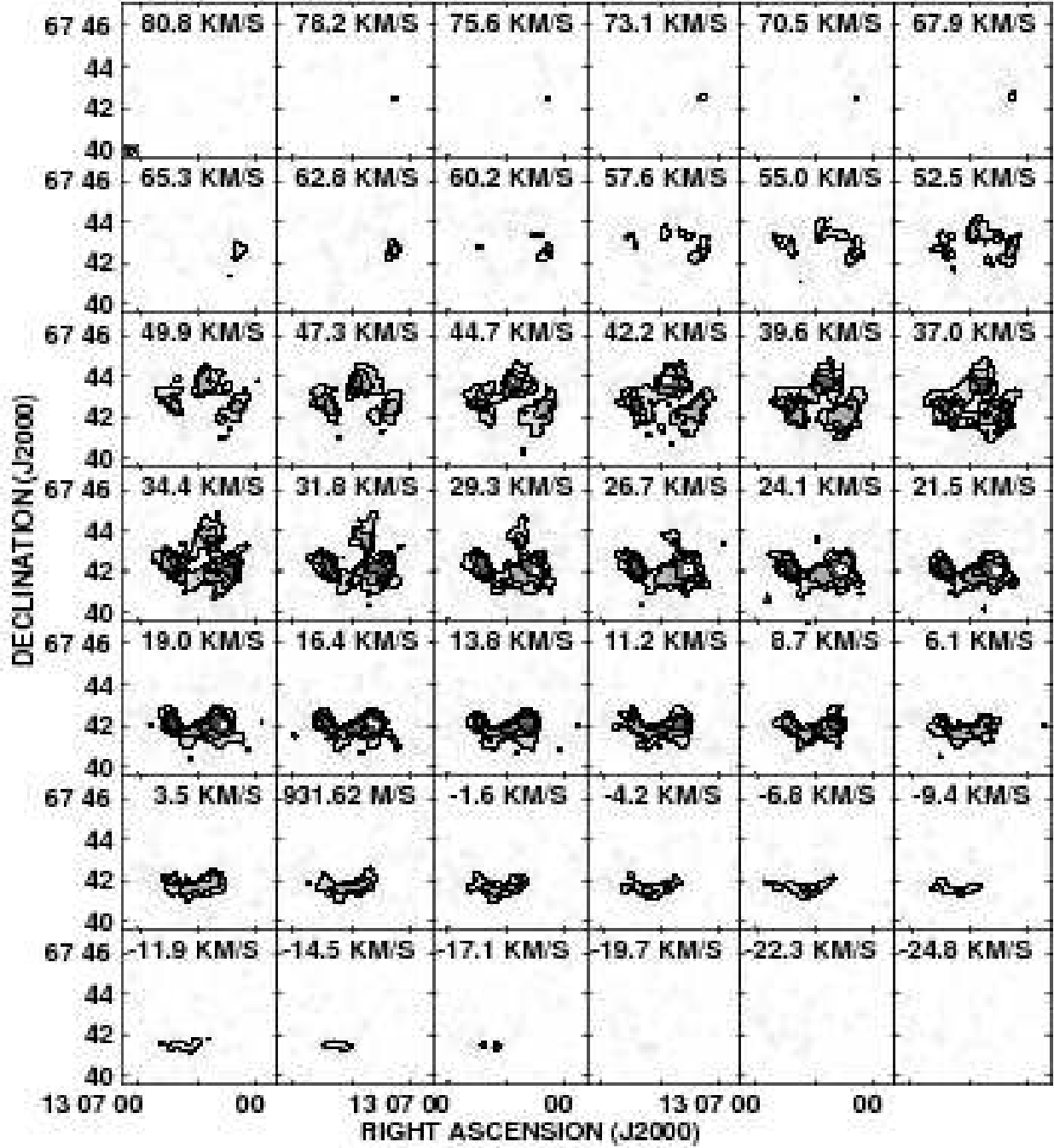}
\epsscale{1.0}
\caption{Channel maps of DDO165 with a circular beam size of
20\arcsec\ $\times$ 20\arcsec\ (indicated in the lower left corner of
the first plane). The contours are at the 4, 8, 16, and 32 $\sigma$
levels ($\sigma$ $=$ 7.1\,$\times$\,10$^{-4}$ Jy\,Bm$^{-1}$).  Note the
lack of \HI\ in the giant \HI\ hole, prominent between $\sim$50 \kms\
and $\sim$25 \kms.}
\label{figcap1}
\end{figure}

\HI\ emission from DDO\,165 is detected in the velocity range $-$20 to
$+$80 \kms.  Figure~\ref{figcap1} presents the individual channel maps
from the 20\arcsec\ $\times$ 20\arcsec\ datacube.  The curious nature
of the \HI\ in DDO\,165, which was previously noted by
\citet{kamphuis96} and \citet{stil02a}, is immediately evident in this
figure.  If solid-body rotation were present (as is typical for dIrrs;
{Skillman 1996}\nocite{skillman96}), it would appear as a smooth
change in the position of neutral gas as a function of velocity
\citep[e.g.,][]{cannon10}.  Ordered rotation is decidedly absent in
DDO\,165.

The channel maps reveal two main components of the \HI\ gas.  Moving
upward from negative velocities, there is emission in nearly all
channels from the ``southern'' \HI\ component, which contains two
column density peaks on the east and west sides of the system
(consider, for example, the panel at 21.5 \kms). The ``northern''
\HI\ component appears between $\sim$ 24 to 58 \kms.  These two
features surround a large cavity in the center of the
\HI\ distribution.  This \HI\ hole is especially evident between
radial velocities of $\sim$50 to 25 \kms.  We discuss this ``hole'' in
detail below.

\begin{figure}
\epsscale{1.0}
\plotone{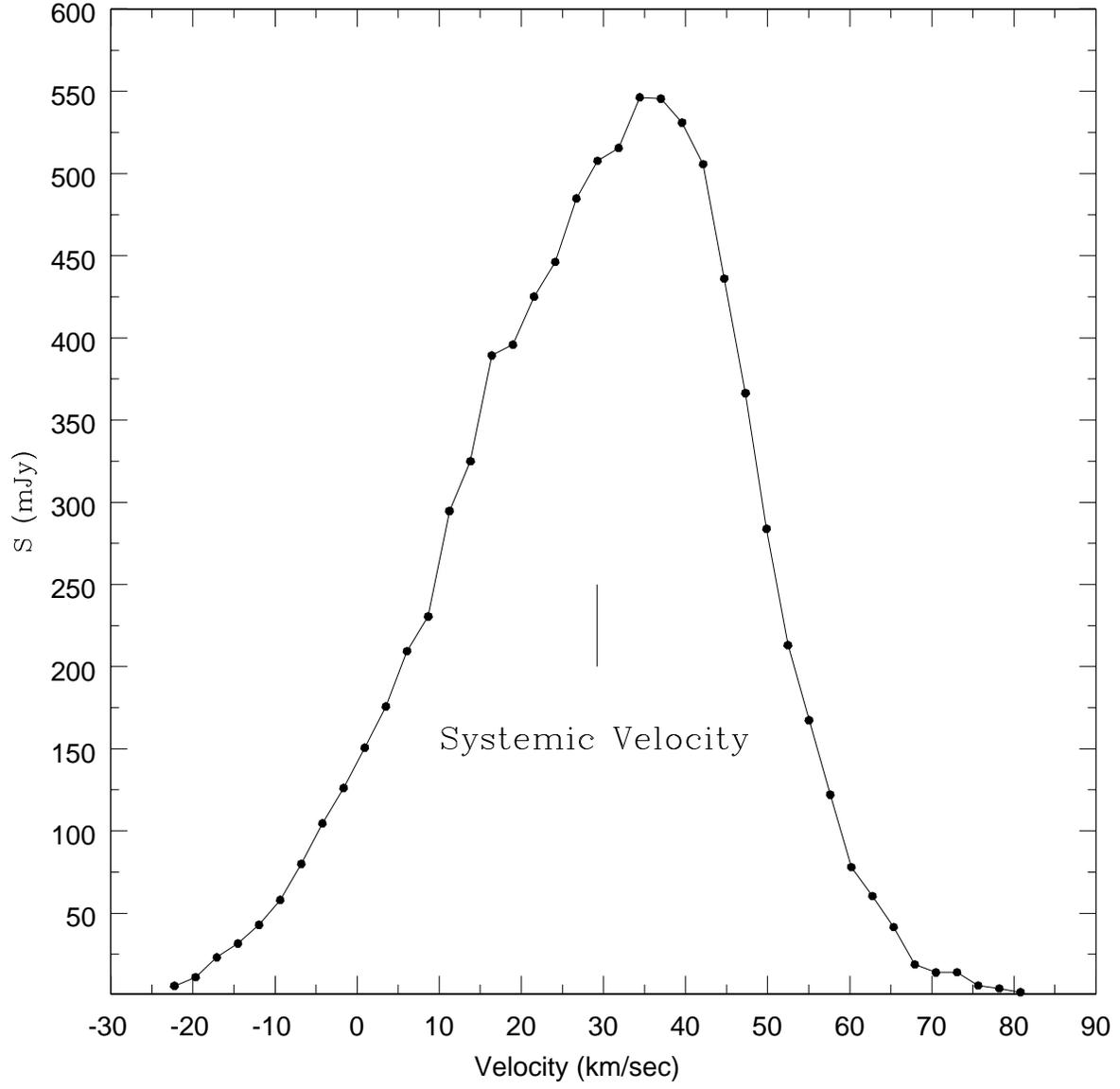}
\epsscale{1.0}
\caption{Global \HI\ profile of DDO\,165, calculated by summing the
flux from each of the channels with real emission in the 20\arcsec\
resolution cube.  The systemic velocity of (29.2\pom2) \kms\ is
derived from this profile.}
\label{figcap2}
\end{figure}

The \HI\ flux from each channel of the 20\arcsec\ $\times$ 20\arcsec\
datacube is plotted as a function of velocity in the global profile
presented in Figure~\ref{figcap2}.  The profile is approximately
Gaussian, although the higher velocity side contains significantly
more flux than the lower velocity side; it is very similar in shape to
the single-dish profile shown in \citet{huchtmeier03}.  The departures
from symmetry are suggestive of the presence of multiple \HI\
components (see discussion above); our kinematic analysis verifies
this interpretation (see \S~\ref{S4}).  The systemic velocity, V$_{\rm
sys}$ $=$ (29\pom2) \kms, is found from the velocity at full width
half maximum of this global profile; at the adopted distance of
4.47\pom0.20 Mpc \citep{karachentsev02}, this low systemic velocity
implies a significant peculiar motion within the M\,81 group.

The total \HI\ flux integral was derived from the 20\arcsec\ cube,
yielding S$_{\rm HI}$ = 23.2\pom2 Jy\,\kms.  This is 18\% lower
than the single dish flux integral of 28.2 Jy \kms\ found by
\citet{huchtmeier03} using Effelsberg 100\,m observations.  We note
that our maximum single-channel flux density of 550\pom50 mJy is in
fair agreement with the corresponding maximum value from
\citet{huchtmeier03} of 601\pom11 mJy. As discussed above, we applied
residual flux scaling in the imaging process, and thus the small
difference between these global flux integrals is most likely due to
the lack of very short baselines in our interferometric observations,
and our corresponding insensitivity to diffuse, extended neutral gas.
The total detected \HI\ mass is (1.1\pom 0.3)$\times$10$^{\rm 8}$
\msun.

\begin{figure}
\epsscale{1.0}
\plotone{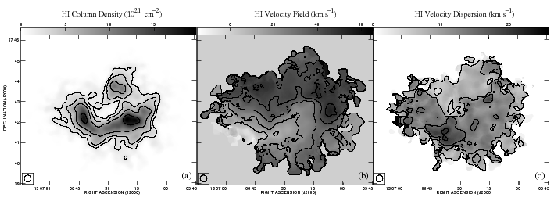}
\epsscale{1.0}
\caption{Moment 0 (a), moment 1 (b), and moment 2 (c) maps of DDO\,165
with a beam size of 20\arcsec $\times$ 20\arcsec\ (indicated in the
lower left corners). The contours in panel (a) are at the (2, 4, 8,
16)\,$\times$\,10$^{20}$ cm$^{-2}$ levels. The contours in panel (b)
are at the (10, 20, 30, 40, 50, 60) \kms\ levels. The contours in
panel (c) are at the (5, 10, 15, 20) \kms\ levels.}
\label{figcap3}
\end{figure}

\begin{figure}
\epsscale{1.0}
\plotone{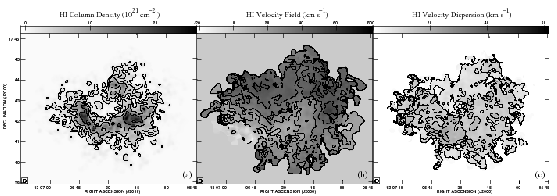}
\epsscale{1.0}
\caption{Moment 0 (a), moment 1 (b), and moment 2 (c) maps of DDO\,165
with a beam size of 10\arcsec $\times$ 10\arcsec\ (indicated in the
lower left corners). The contours in panel (a) are at the (2, 4, 8,
16)\,$\times$\,10$^{20}$ cm$^{-2}$ levels. The contours in panel (b)
are at the (10, 20, 30, 40, 50, 60) \kms\ levels. The contours in
panel (c) are at the (5, 10, 15, 20) \kms\ levels.}
\label{figcap4}
\end{figure}

Figures~\ref{figcap3} and \ref{figcap4} show moment zero (representing
\HI\ surface density), moment one (representing intensity-weighted
velocity field), and moment two (representing velocity dispersion)
images at 20\arcsec\ and 10\arcsec\ resolution, respectively.  The
\HI\ column density peaks in three regions in DDO\,165; the two largest
peaks (N$_{\rm HI}$ $\simeq$2\,$\times$10$^{21}$ cm$^{-2}$) are
associated with the southern component of the system, which comprises
$\sim$85\% of the total \HI\ mass.  The northern component reaches a
maximum column density of N$_{\rm HI}$
$\simeq$1.1\,$\times$10$^{21}$.  Low column density gas connects the
northern and southern features, primarily in the western section of
the disk.  We estimate the mass of the northern component to be
(1.6\,\pom\,0.4)\,$\times$\,10$^7$ \msun.

A central \HI\ depression is very prominent in the moment zero images.
The physical diameter of this feature is $\sim$2.2\,$\times$\,1.1 kpc
(see further discussion below), making this one of the largest such
known in a dwarf galaxy.  We hereafter refer to this feature as the
central \HI\ hole.  The size and morphology is reminiscent of the
central holes in several well-studied systems (e.g., M81\,dwA,
{Sargent \etal\ 1983}\nocite{sargent83}; Sextans\,A, {Skillman
  \etal\ 1988}\nocite{skillman88}; Holmberg\,I, {Ott
  \etal\ 2001}\nocite{ott01}). Note that there is only very low column
density neutral gas in the northeast region of the galaxy.  In that
regard, the central hole can be considered incomplete, as it is not
surrounded by higher column density gas around the entire
circumference.

As expected from the channel maps (see Figure~\ref{figcap1}), the
representations of the velocity field of DDO\,165 shown in
Figures~\ref{figcap3} and \ref{figcap4} are very complex and highly
disturbed.  There is evidence for rotation of the southern component,
although the axis of rotation is position dependent.  In the southeast
region, the isovelocity contours suggest rotation on a position angle
of $\sim$310\degr; however, the central region of the southern
component appears to be rotating more or less east-west.  The northern
component shows weak evidence for rotation at a third position angle
($\sim$0\degr).  The velocity field of DDO\,165 is highly irregular
compared to most dwarf galaxies of comparable mass \citep[see,
  e.g.,][]{begum08}; we explore the velocity field further in
\S~\ref{S4}.

We attempted a tilted ring analysis of these velocity fields using the
GIPSY\footnote{The Groningen Image Processing System (GIPSY) is
  distributed by the Kapteyn Astronomical Institute, Groningen,
  Netherlands.} task ROTCUR.  We experimented with numerous iterations
of the fit parameters, including dynamical center position,
inclination, and rotation position angle.  Since ROTCUR assumes a
symmetric velocity distribution for its input, it is perhaps not
surprising that our fits did not identify a unique set of parameters
to describe the complex velocity field of DDO\,165.  The primary
difficulty is in identifying a unique dynamical center position;
holding other parameters constant, we were able to derive a convergent
solution for essentially any central position.  In a final attempt to
attain a successful solution, we convolved the velocity field to low
resolution (1\arcmin), but were again unable to derive an unambiguous
rotation curve.

Previous analysis of the \HI\ kinematics by \citet{stil02a} identified
similar difficulties in fitting the \HI\ velocity field.  The position
angle of the maximum velocity gradient from that work was found to be
(130\degr\pom10\degr), on a line from southeast to northwest
(equivalent to [310\degr\pom10\degr] measured east of north).  In our
tilted ring analysis, we identify the same position angle for the
largest coherent velocity gradient.  We again stress that a unique
rotation curve was not identified, but nonetheless use this position
angle in the position-velocity (PV) analysis presented in
\S~\ref{S4.1}.  We note that no unique optical position angle is
listed in the Catalog of Principal Galaxies \citep{paturel03} or the
Sloan Digitized Sky Survey \citep{abazajian09}.

The velocity dispersion images shown in Figures~\ref{figcap3} and
\ref{figcap4} reveal interesting characteristics of the \HI\
kinematics.  The southern component has a higher average velocity
dispersion ($\sim$12 \kms) than the northern component ($\sim$8 \kms);
the difference is larger than the velocity resolution of the data
(2.56 \kms).  The $\sim$12 \kms\ average in the southern component is
slightly larger than the characteristic $\sigma_{\rm V}$ values found
for three {\it THINGS} dwarf galaxies in \citet{tamburro09}.
Interestingly, the characteristic $\sigma_{\rm V}$ value for the
northern component is in good agreement with that average.  Note that
$\sigma_{\rm V}$ is elevated to values larger than 20 \kms\ in the
center of the southern component.  This region of increased velocity
dispersion is located between the two column density peaks.  It is
coincident with the position at which the isovelocity contours show
significant changes in direction (see also discussion above).  Taken
as a whole, these interesting characteristics of the \HI\ morphology
and dynamics prompted a detailed investigation of the small-scale gas
motions within DDO\,165; we discuss the results of those analyses in
\S~\ref{S4}.

\subsection{Comparing the Stellar and Gaseous Components}
\label{S3.2}

\begin{figure}
\epsscale{1.0}
\plotone{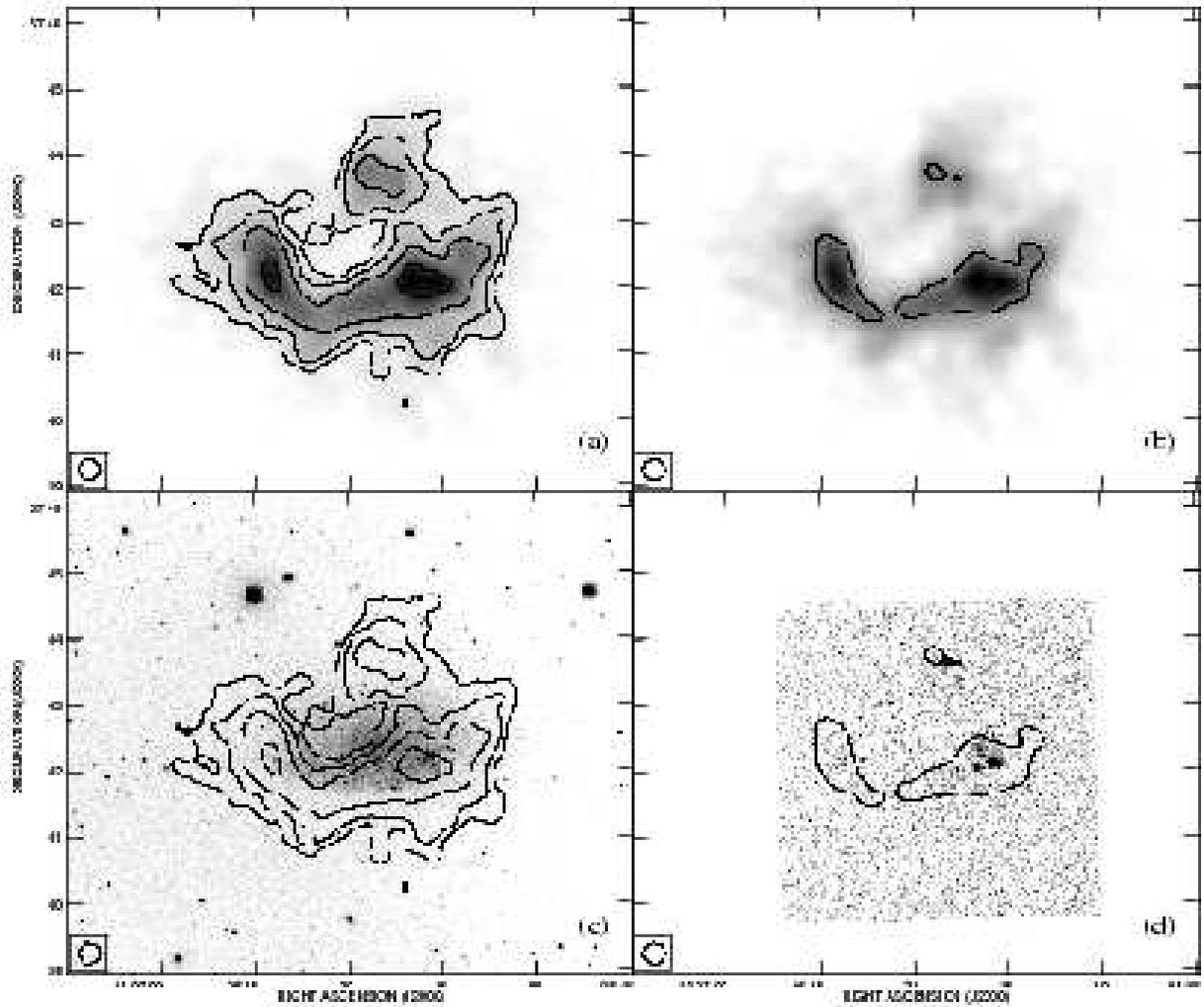}
\epsscale{1.0}
\caption{Multiwavelength images of DDO\,165.  Panels (a) and (b) show
the same 20\arcsec\ resolution \HI\ column density map as
Figure~\ref{figcap3}(a); the contours in (a) are at the (2, 4, 8,
16)\,$\times$\,10$^{20}$ cm$^{-2}$ levels, while the contour in (b) is
at the 10$^{21}$ cm$^{-2}$ level.  Panel (c) shows the R-band image
from the KPNO 2.1\,m telescope, overlaid with the same contours as
panel (a).  Panel (d) shows the continuum-subtracted \halpha\ image
from the KPNO Bok telescope, overlaid with the same contours as panel
(b).  Note the correspondence between high column density \HI\ gas and
the locations of ongoing SF.}
\label{figcap5}
\end{figure}

In Figure~\ref{figcap5}, we compare the neutral gas distribution with
optical observations.  Figures~\ref{figcap5}(a) and (b) show the same
moment zero image as Figure~\ref{figcap3}.  The contours in
Figure~\ref{figcap5}(a) are the same as in Figure~\ref{figcap3}; the
contour level in Figure~\ref{figcap5}(b), used below and throughout
Paper~II, is at the 10$^{21}$ cm$^{-2}$ level, representing the
canonical SF threshold (Kennicutt
1989\nocite{kennicutt89}, 1998\nocite{kennicutt98}).  Note that all
three aforementioned column density peaks exceed this surface density
level.  Figure~\ref{figcap5}(c) compares the R-band image with the
\HI; the contours are the same as in Figure~\ref{figcap5}(a).  The
high surface brightness optical component is not centered on the \HI\
hole; note the dramatic drop in stellar surface brightness in the
southern component.  Interestingly, this boundary is coincident with
substantial \HI\ column densities (N$_{\rm HI}$ $>$
8\,$\times$\,10$^{20}$ cm$^{-2}$).  The northern \HI\ component is
significantly offset from the main optical body of DDO\,165.

Figure~\ref{figcap5}(d) compares the locations of active SF as traced
by \halpha\ emission with the locations of high \HI\ surface density;
the contour is again at the 10$^{21}$ cm$^{-2}$ level.  Each
\HI\ surface density maximum has associated \halpha\ emission.  We
find 10 discrete \halpha\ regions, and catalog their coordinates and
flux measurements in Table~\ref{t3}.  The total flux from these 10
regions is 8.5 $\times$ 10$^{-14}$ erg\,s$^{-1}$\,cm$^{-2}$. Our total
\halpha\ flux agrees with the global value presented by \citet{lee09}
at the $\sim$30\% level; we attribute the difference to diffuse
\halpha\ flux not enclosed in the 10 discrete \halpha\ regions
summarized in Table~\ref{t3}.

\begin{figure}
\epsscale{1.0}
\plotone{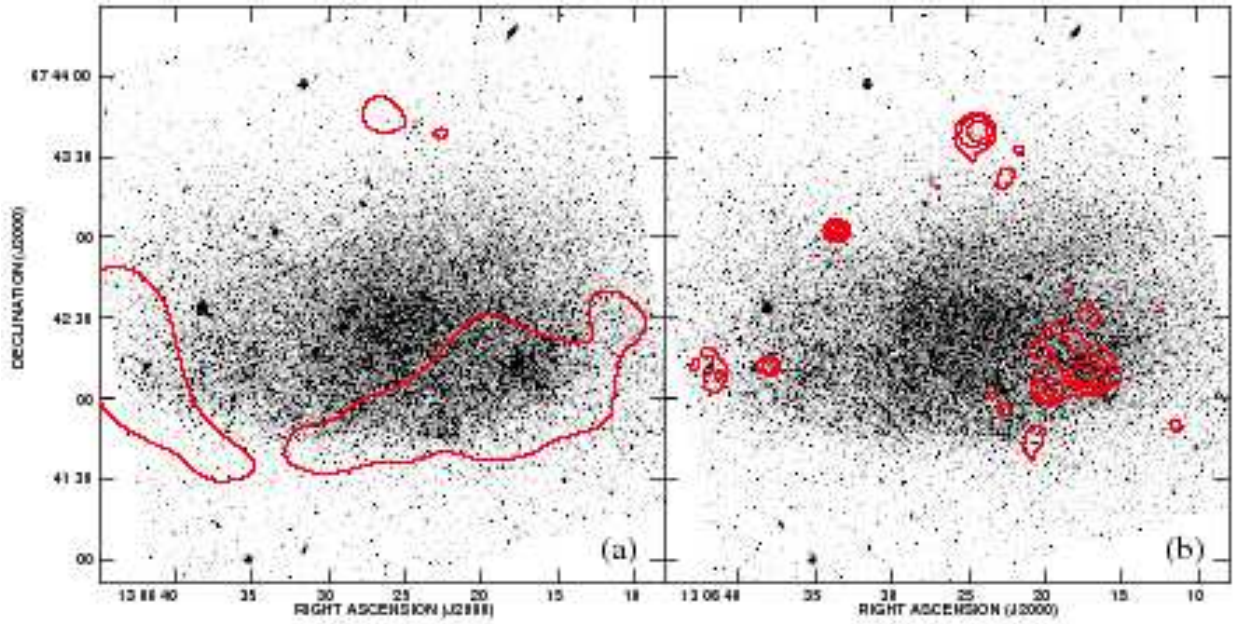}
\epsscale{1.0}
\caption{{\it HST}/ACS F555W images of DDO\,165 with \HI\ (a) and
\halpha\ (b) contours overlaid.  The \HI\ contour in panel (a) is at
the 10$^{21}$ cm$^{-2}$ level; the \halpha\ contours in (b) are at the
(2, 4, 8, 16)\,$\times$\,10$^{-18}$ erg\,s$^{-1}$\,cm$^{-2}$ levels.
Note that there are three actively star-forming regions in DDO\,165
(the compact \halpha\ source in the northeast is an imperfectly
subtracted foreground star), and that each is associated with high
column density neutral gas.}
\label{figcap6}
\end{figure}

Figure~\ref{figcap6} presents a comparison of the \HI\ and
\halpha\ emission with the {\it HST} V-band image (see also the
detailed discussion in Paper~II).  The \HI\ contour is at the
10$^{21}$ cm$^{-2}$ level, while the \halpha\ contours are at levels
of (2,4,8,16)\,$\times$\,10$^{-18}$ erg\,s$^{-1}$\,cm$^{-2}$.  Note by
comparison with Figure~\ref{figcap5} that the large central \HI\ hole
is offset from the center of the stellar distribution. The truncated
stellar population in the southern region is again evident; the
\HI\ surface density is highest along this feature.  Ongoing SF is
associated with the largest stellar association in the southwest.  The
\halpha\ emission in the southeast is apparently not associated with
high surface brightness stellar clusters and likely arises from
individual OB stars.

\begin{figure}
\epsscale{1.0}
\plotone{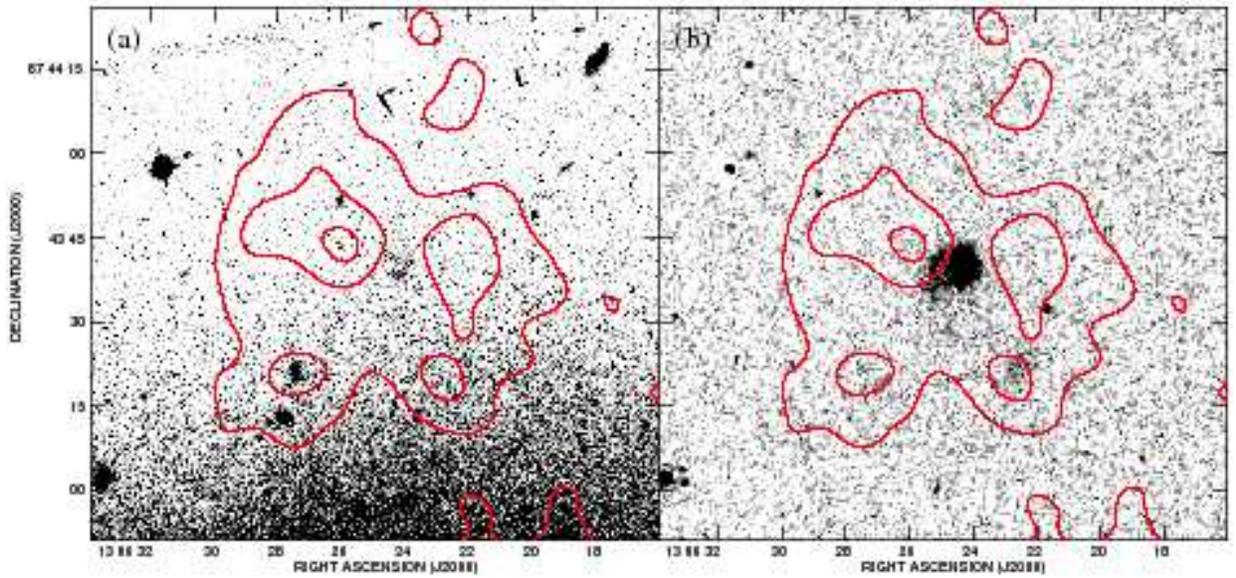}
\epsscale{1.0}
\caption{Zoomed optical images of the northern \HI\ cloud and
associated star forming region.  Panel (a) shows the {\it HST}/ACS
F555W image and panel (b) shows the continuum-subtracted \halpha\
image.  The contours show \HI\ column densities of (5, 10,
15)\,$\times$\,10$^{20}$ cm$^{-2}$ at 10\arcsec\ resolution.  Note
that the optical stellar association and \halpha\ emission is located
immediately on the edge of the highest \HI\ column density peak in
this northern region.  Fainter \halpha\ emission is also associated
with two of the smaller \HI\ local maxima.}
\label{figcap7}
\end{figure}

The nature of SF in the northern region of DDO\,165 is very
interesting.  The stellar density maps in \citet{weisz08} show a
smooth decrease in density of both red and blue stars from the central
maximum to the northern region.  A small over-density is apparent at
the location of the northern \HI\ component in the blue stellar
density map.  Yet, the northern \HI\ cloud has dense neutral gas above
the 10$^{21}$ cm$^{-2}$ level as well as a compact \halpha\ source.
To further investigate this curious northern component,
Figure~\ref{figcap7} shows a closer view of the {\it HST} and \halpha\
images of this region, overlaid with contours from the 10\arcsec\
moment zero image (see Figure~\ref{figcap4}).  This zoomed field
clearly shows a stellar association coincident with the \halpha\
maximum, nestled between two \HI\ surface density peaks.  Further, the
very diffuse \halpha\ emission features seen at (13:06:22.5, 67:43:20)
and (13:06:27.5, 67:43:20) are also associated with local \HI\
over-densities that are smoothed out in the lower resolution surface
density images.

\section{Kinematic Analysis of the HI Gas}
\label{S4}

DDO\,165 displays numerous abnormal characteristics.  Its velocity
field is highly confused, without obvious signatures of coherent
global rotation.  Its stellar distribution is abruptly truncated in
the southern region, where the bulk of the neutral gas mass and
ongoing SF are located.  A region of ongoing SF is located well
outside of the main optical body, cospatial with surprisingly dense
neutral gas in the northern \HI\ component.  A kpc-scale hole
dominates the morphology of the neutral gas. We now seek to better
understand the origins of these enigmatic features in DDO\,165 using
the kinematic information available in the neutral gas.

\subsection{Spatially Resolved Position-Velocity Diagrams}
\label{S4.1}

\begin{figure}
\epsscale{0.8}
\plotone{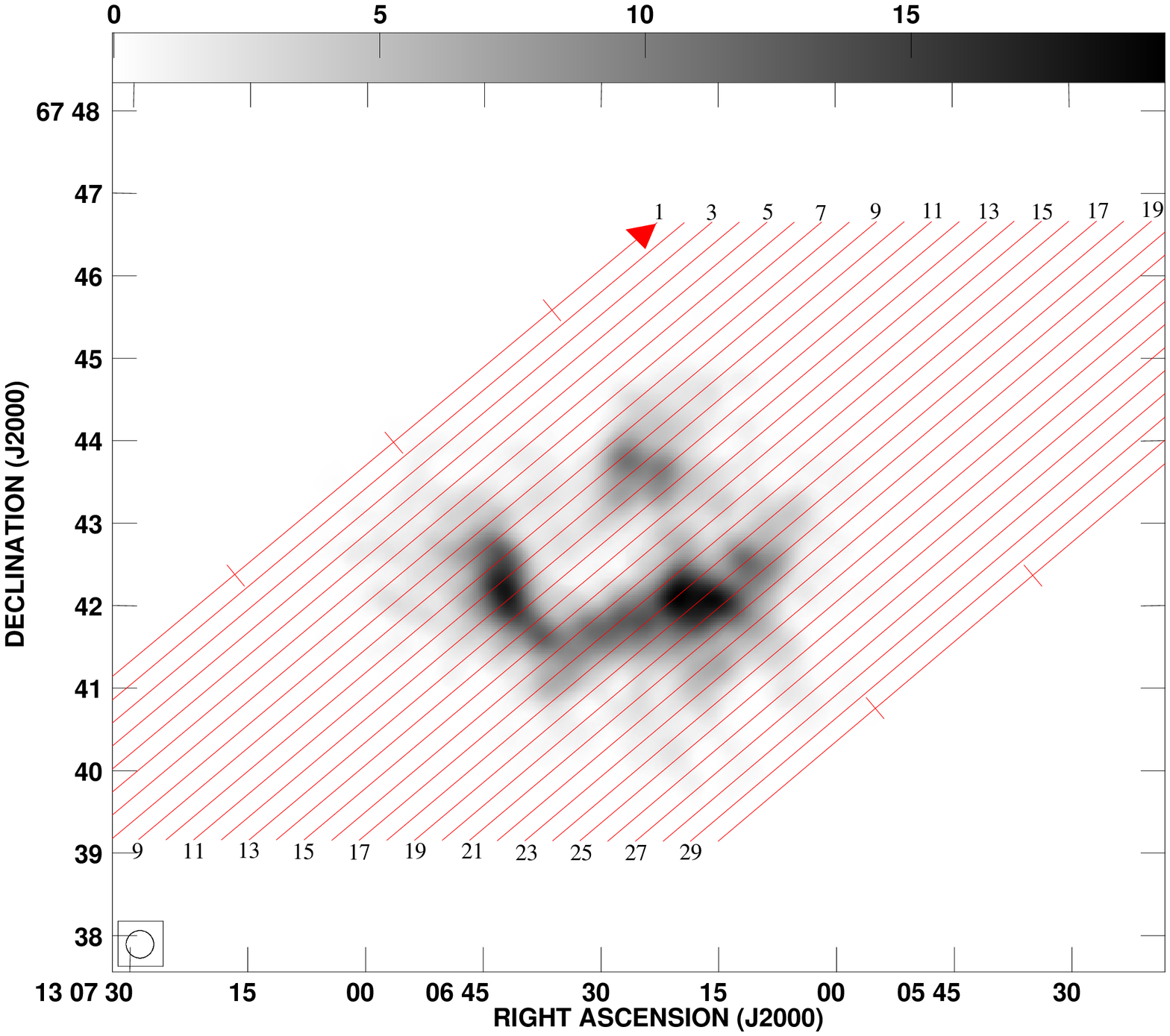}
\epsscale{1.0}
\caption{20\arcsec\ resolution moment 0 map of DDO\,165 overlaid with
major-axis (position angle of 310\degr, measured east of north) PV
slice positions. Each slice has a width of, and is separated by,
20\arcsec\ (the beam width).  The resulting PV diagrams from each
slice are shown in Figure~\ref{figcap10}.  The arrow on slice \#1
indicates the direction of the slices (the same arrow can be seen on
the bottom of each slice in Figure~\ref{figcap10}). The tick marks
along slices \#1 and \#30 indicate distances of 150\arcsec,
300\arcsec, 450\arcsec, and 600\arcsec, moving along a slice from the
bottom in the direction of the arrow.  The color wedge shows the 
\HI\ column density scale in units of 10$^{20}$ cm$^{-2}$.}
\label{figcap8}
\end{figure}

\begin{figure}
\epsscale{0.8}
\plotone{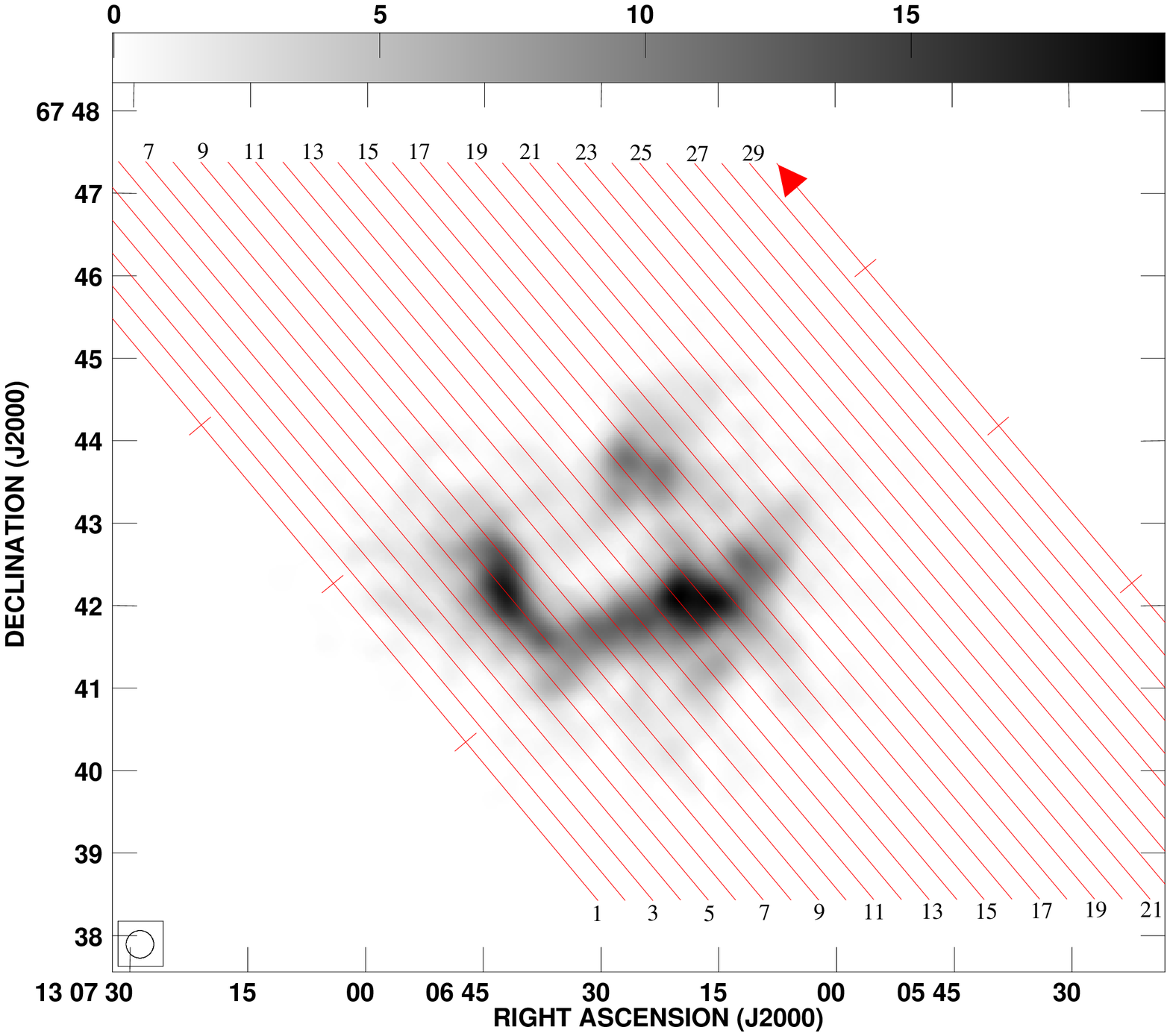}
\epsscale{1.0}
\caption{20\arcsec\ resolution moment 0 map of DDO\,165 overlaid with
minor-axis (position angle of 40\degr, measured east of north) PV
slice positions. Each slice has a width of, and is separated by,
20\arcsec\ (the beam width).  The resulting PV diagrams from each
slice are shown in Figure~\ref{figcap11}.  The arrow on slice \#30
indicates the direction of the slices (the same arrow can be seen on
the bottom of each slice in Figure~\ref{figcap11}). The tick marks
along slices \#1 and \#30 indicate distances of 150\arcsec,
300\arcsec, 450\arcsec, and 600\arcsec, moving along a slice from the
bottom in the direction of the arrow.  The color wedge shows the 
\HI\ column density scale in units of 10$^{20}$ cm$^{-2}$.}
\label{figcap9}
\end{figure}

The unusual \HI\ properties of DDO\,165 led us to first analyze the
global neutral gas kinematics. We thus began by using GIPSY/SLICEVIEW
to create 60 PV slices through the 20\arcsec\ datacube.  30 slices are
taken along the major axis (310\degr\,\pom10\degr; see \S~\ref{S3.1}),
and 30 are taken 90\degr\ from the major axis.  These slices are the
width of, and separated by, the low-resolution beam size (20\arcsec).
Figures~\ref{figcap8} and \ref{figcap9} show the slice locations
overlaid on the 20\arcsec\ moment zero map for PV cuts at position
angles of 310\degr\ and at 40\degr, respectively.  Note that two
adjacent PV cuts overlap by half of the beam width; thus every other
PV cut is independent.  By creating the slices with a width of and
separated by the beam size, we can study the kinematics throughout the
entire galaxy and minimize selection effects caused by arbitrary slice
locations.

\begin{figure}
\epsscale{1.0}
\plotone{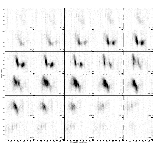}
\epsscale{1.0}
\caption{Mosaic of the major-axis PV slices indicated in
Figure~\ref{figcap8}. Each panel represents a 20\arcsec\ wide PV cut
through the 20\arcsec\ resolution \HI\ data cube.}
\label{figcap10}
\end{figure}

\begin{figure}
\epsscale{1.0}
\plotone{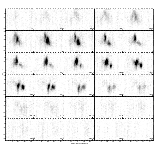}
\epsscale{1.0}
\caption{Mosaic of the minor-axis PV slices indicated in
Figure~\ref{figcap9}. Each panel represents a 20\arcsec\ wide PV cut
through the 20\arcsec\ resolution \HI\ data cube.}
\label{figcap11}
\end{figure}

The PV diagrams along each of these slices are shown in
Figures~\ref{figcap10} and \ref{figcap11} for the major and minor
axes, respectively.  To aide interpretation, directional arrows are
shown in both the locations of the slices (Figures~\ref{figcap8} and
\ref{figcap9}) and in the resulting PV diagrams
(Figures~\ref{figcap10} and \ref{figcap11}).  Moving from southeast to northwest
along the slices shown in Figure~\ref{figcap8} corresponds to moving
from left to right in a given panel of Figure~\ref{figcap10}.
Similarly, moving from southwest to northeast along the slices shown in
Figure~\ref{figcap9} corresponds to moving from left to right in a
given panel of Figure~\ref{figcap11}.  The tick marks along the first
and last slices in Figures~\ref{figcap8} and \ref{figcap9} are at the
150\arcsec, 300\arcsec, 450\arcsec, and 600\arcsec\ positions,
corresponding to the same distance markers on the abscissae of
Figures~\ref{figcap10} and \ref{figcap11}.

\subsubsection{Kinematically Distinct HI Components}
\label{S4.1.1}

The major axis PV slices (Figures~\ref{figcap8} and \ref{figcap10})
immediately demonstrate that the northern and southern \HI\ components
are kinematically distinct.  Moving from slice\,8 to slice\,13, the
two \HI\ components appear at velocities in common ($\sim$30--60 \kms,
though the southern component subtends a much larger total velocity
range) but are separated by physical distances of 1 kpc or more.  The
region between these two features is easily identified as the giant
\HI\ hole in the center of the galaxy; slices\,9 through 15 clearly
show the evacuated nature of the central hole and the lack of high
surface brightness \HI\ emission from within it.

The kinematic discontinuity between the northern and southern region
is confirmed in the minor-axis PV cuts shown in Figures~\ref{figcap9}
and \ref{figcap11}.  Note that the cuts that pass through the northern
\HI\ component in Figure~\ref{figcap9} also pass through gas in the
western region of the southern \HI\ component; however, the cuts that
pass through the northern \HI\ component in Figure~\ref{figcap8} pass
through gas in the eastern region of the southern \HI\ component.
While the major-axis cuts that pass through the northern component
cross the giant \HI\ hole, the minor axis cuts that pass through the
northern component do not.  And yet, remarkably, slices\,13 through 17
in Figure~\ref{figcap11} clearly show that the northern and southern
\HI\ components are separated physically and without \HI\ emission
connecting them in velocity space. 

Kinematically distinct \HI\ features have been observed in other
nearby dwarf galaxies, including NGC 1705 \citep{meurer98}, NGC 1569
\citep{stil02b}, NGC 625 \citep{cannon04}, and NGC 5253
\citep{kobulnicky08}.  It is interesting to note that each of these
systems is a well-studied, ``canonical'' dwarf starburst galaxy,
typically with multi-phase outflows and high equivalent width \halpha\
emission from ongoing SF.  In contrast, DDO\,165 shows only very weak
ongoing SF.  However, the intensity of SF was greater in the past, as 
established in \citet{weisz08}, \citet{lee09}, and \citet{mcquinn10b}.
We discuss the recent SF history in detail in Paper~II.  

We note that based on the \HI\ kinematics alone, we are able to
exclude certain origins for the kinematically distinct
\HI\ components.  Since the components have velocities in common but
are separated by more than a kpc, it is unlikely that two superposed
or counter-rotating disks can explain the features seen in the PV
diagrams.  Similarly, a coherently rotating disk (common in dIrr
galaxies) is clearly excluded.  Finally, while the ISM is clearly
highly disturbed (see also the discussion in \S~\ref{S4.1.2} below),
recent simulations have shown that most of the energy from turbulence 
is contained on scales of a few hundred pc or less \citep{joung06};
it is unlikely that turbulence could be capable of remaining coherent
over kpc scales or of injecting sufficient energy to cause bulk motion
of the $\sim$10$^7$ \msun\ of neutral gas associated with the northern
\HI\ component. 

The \HI\ kinematics alone do not, however, allow us to differentiate
between an infall or an outflow (i.e., stellar feedback) origin for
the kinematic discontinuity.  The similar velocities of the two \HI\
components is the primary difficulty in differentiating between these
two scenarios using our \HI\ data.  If the northern component has been
ejected in a massive blowout event, then we might expect \HI\ at
velocities intermediate between the two components; such gas is not
seen in our PV diagrams.  Similarly, if the northern cloud is falling
into the main disk of DDO 165, we might expect tenuous \HI\ gas
extended behind it, both spatially and in velocity space (see, e.g.,
{Hunter \etal\ 1998}\nocite{hunter98b}).  After smoothing our data to
a 90\arcsec\ $\times$ 90\arcsec\ beam size, no such extended structure
is found.  Further, the infall hypothesis has difficulty with the
evacuated giant \HI\ hole, which would lie ahead of the motion of the
northern component in an infall model.

We conclude that simple outflow and infall hypotheses have
inconsistencies with our \HI\ data, but that both remain viable
mechanisms for the creation of the two \HI\ components of DDO\,165.
We are mindful that relative motion between the two \HI\ components
remains possible if the entire DDO\,165 system is seen face-on.
Without additional avenues with which to exclude the infall hypothesis
based on \HI\ kinematics, we interpret the nature of the kinematically
distinct northern component in the context of the past SF history and
the blowout paradigm in Paper~II.

\subsubsection{Localized High-Velocity HI Features}
\label{S4.1.2}

The spatially resolved PV slices through the 20\arcsec\ datacube
reveal a remarkable wealth of localized, high-velocity neutral gas
features in the disk of DDO\,165.  These features are spatially
compact but extended in velocity space; such features have been
observed in a few starbursting dwarf galaxies (e.g., NGC\,625 -
{Cannon \etal\ 2004}\nocite{cannon04}; NGC\,5253 - {Kobulnicky \&
  Skillman 2008}\nocite{kobulnicky08}).  Examining the major-axis PV
slices (Figures~\ref{figcap8} and \ref{figcap10}) highlights
high-velocity features in the southeast region of the
\HI\ distribution.  Slices 8 through 15 cross the southeast
\HI\ column density peak, the giant \HI\ hole, and the northern
\HI\ component.  Note that the velocity extent of the \HI\ gas in the
southeast region spans more than 50 \kms\ in each of these slices,
from $\sim$0 \kms\ through $\sim$50 \kms.  The 20\arcsec\ slice width
spans only $\sim$430 pc at the adopted distance.

The high surface density neutral gas in the southwest region is
extended along the PV slice major axis, and is thus more easily
studied in the minor-axis slices shown in Figures~\ref{figcap9} and
\ref{figcap11}.  There, slices 9 through 11 sample the center of the
southern \HI\ component before traversing the giant \HI\ hole.  While
the velocity extent of these \HI\ features is slightly smaller than
the southeast features discussed above, they still span more than
$\sim$40 \kms, from $\sim$$-$10 \kms\ through $\sim$30 \kms.
Continuing westward along the southern component, slices 13 through 15
pass through the main \HI\ column density peak and its associated
active SF region.  The $\sim$40 \kms\ velocity extent of the neutral
gas is again seen in the southern component.  Finally, moving from
slice 16 through slice 18, the material in the western disk is also
extended in velocity space.  These localized features span up to
$\sim$50 \kms, from $\sim$20 \kms\ through $\sim$70 \kms\ (see
especially slice 17).

The physical conditions of the gas in the southern and northern
components appear to be quite different.  Most of the gas in the
southern component is extended in velocity space (confirming the
elevated velocity dispersion discussed in \S~\ref{S3.1}).  However,
all slices through the northern \HI\ component (slices 9 through 14 in
the major-axis PV diagrams shown in Figures~\ref{figcap8} and
\ref{figcap10}, and slices 13 though 18 in the minor-axis PV diagrams
shown in Figures~\ref{figcap9} and \ref{figcap11}) show it to be 
comparatively compact in velocity space, spanning $\sim$20-25 \kms\ 
along both axes. 

The information in these spatially resolved PV diagrams, and in the
velocity dispersion images discussed above, highlight a factor of
$\sim$2 difference in velocity extent of localized gas features
between the northern and southern \HI\ components.  Recall by
examining Figures~\ref{figcap5} and \ref{figcap6} that the southern
\HI\ region contains the two largest \HI\ column density maxima, the
most luminous \halpha\ emission source, and a much higher stellar
density than the northern region.  These high velocity gas features
suggest that dynamical effects, either localized due to feedback from
recent SF or globally due to an interaction, are having a dramatic
impact on the neutral gas in the southern region of DDO\,165.  We
quantitatively interpret the nature of these high velocity features in
the context of the feedback model in Paper~II.

\subsection{Kinematic Search for Coherent \HI\ Holes and Shells in the ISM}
\label{S4.2}

As discussed in \S~\ref{S1}, feedback from massive star evolution can
directly shape the surrounding ISM.  The post-starburst
characteristics, disrupted velocity field, and extended velocity
features in the neutral ISM of DDO\,165 suggest that this feedback
mechanism could be especially efficient at creating ISM structures.
In a simplistic case, one would expect multiple holes and shells in
the neutral gas if the past SF has created coherent structures via
energy deposition in stellar winds and/or SNe, and if those structures
had not yet been erased by either dynamical processes or by the
re-establishment of pressure equilibrium with the surrounding
multi-phase ISM.  We thus undertook a search for such structures in
both PV space and radius-velocity (RV) space.  We use the KPVSLICE and
KSHELL tools in the KARMA\footnote{The KARMA visualizations software
package was developed by Richard Gooch of the Australia Telescope
National Facility (ATNF).} visualization software package, as well as
GIPSY/SLICEVIEW.  Note that each type of analysis has certain
advantages, but the primary gain in moving from PV to RV space is the
increased signal-to-noise ratio as one averages over more pixels than
in a PV slice.

In this kinematic exploration, we characterize potential coherent
structures using the formalism described by \citet{brinks86}.  \HI\
holes and shells are classified into three groups.  In a Type~1 hole,
\HI\ gas has been completely evacuated from a particular region, and
no signature of expansion is apparent in either PV or RV space.  In a
Type~2 feature, one side of the hole has broken out of the plane of
the galaxy, while the other side retains a shell structure.  These
Type~2 structures are perhaps the most difficult to identify; in PV
space they appear as broken ellipses (see also the numerical
simulations of these structures discussed by {Silich \etal\
1996}\nocite{silich96}).  Finally, a Type~3 hole represents a
complete, expanding shell that is easily identified in RV space as an
unbroken half-ellipse or in PV space as a complete ellipse.  Type~3
holes provide the most kinematic information, as both a diameter and
an expansion velocity are directly measurable in RV space.

\begin{figure}
\epsscale{1.0}
\plotone{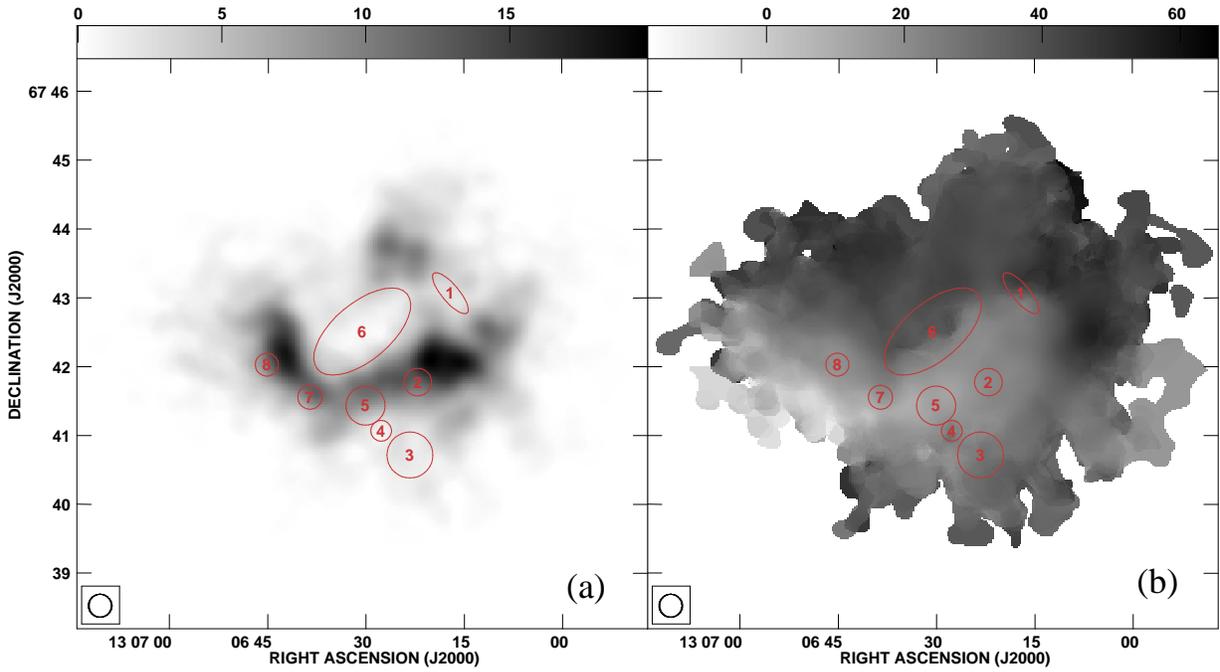}
\epsscale{1.0}
\caption{\HI\ column density image in units of 10$^{20}$ cm$^{-2}$ (a)
and velocity field in units of \kms\ (b) at 20\arcsec\ resolution,
overlaid with the positions of the 8 \HI\ holes identified in our
kinematic analysis.  }
\label{figcap12}
\end{figure}

\begin{figure}
\epsscale{1.0}
\plotone{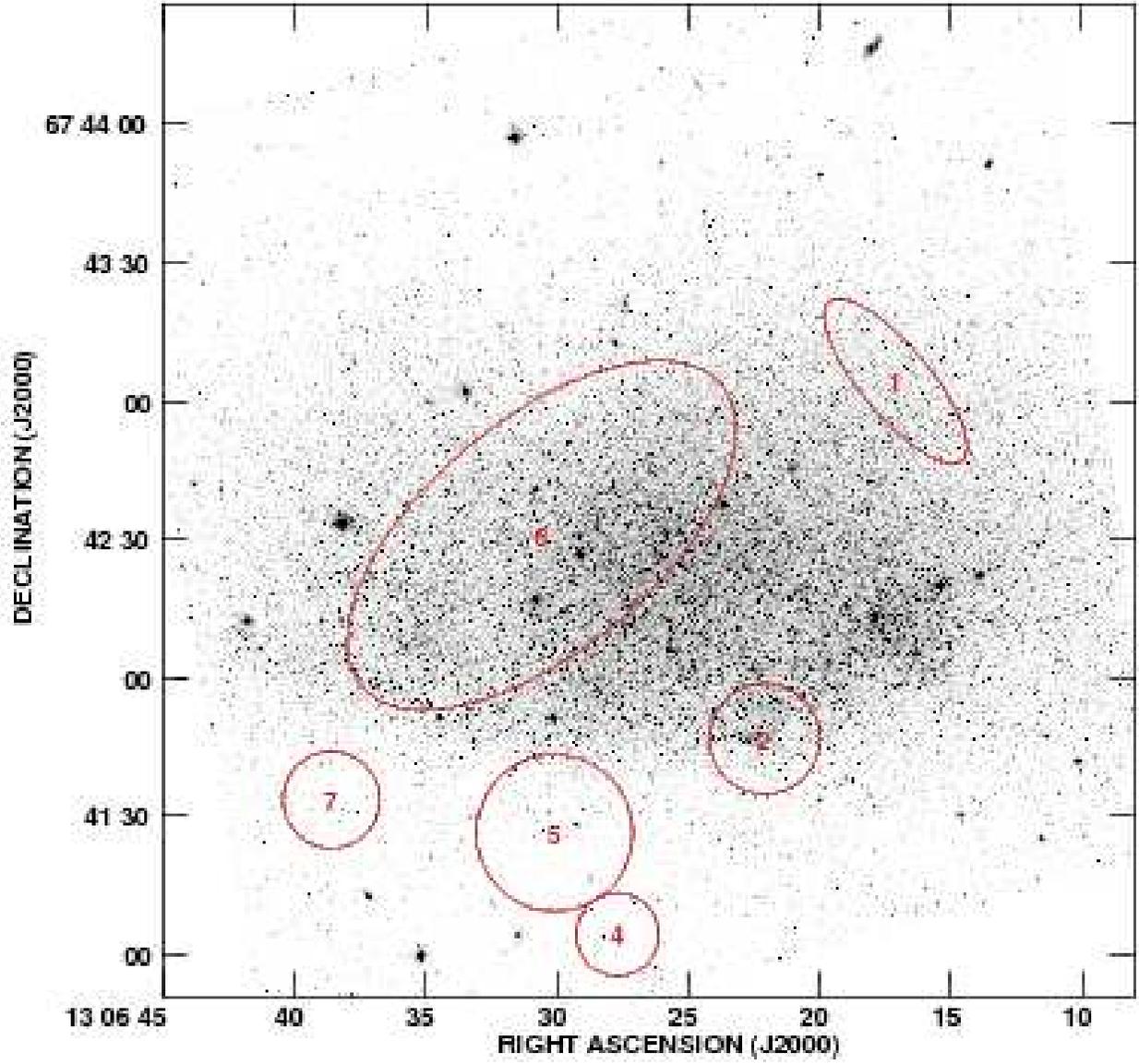}
\epsscale{1.0}
\caption{{\it HST}/ACS F555W images of DDO\,165 with an overlay of the
positions and sizes of the size \HI\ holes that fall completely within
the field of view.  Note that Holes \#3 and \#8 in
Figure~\ref{figcap12} are not covered completely by the {\it HST} images.}
\label{figcap13}
\end{figure}

We searched for these types of structures in both the naturally
weighted cubes (15\arcsec\ and 20\arcsec, sensitive to large-scale
holes and shells) and in the robust weighted cubes (7.5\arcsec\ and
10\arcsec, sensitive to smaller holes and shells).  We examined PV
cuts across the entire system (see discussion above and
Figures~\ref{figcap10} and \ref{figcap11}) and on smaller scales and
at arbitrary position angles.  We also examined RV diagrams with a
minimum radius of the beam size throughout the galaxy. We identified 8
holes and shells in the neutral ISM of DDO\,165 in this analysis; the
locations and properties of these holes are listed in Table~\ref{t4}
in order of increasing central Right Ascension position.  Their
positions, sizes, and orientations are shown overlaid on the
20\arcsec\ \HI\ column density image and velocity field in
Figure~\ref{figcap12}.  Forunately, 6 of these 8 structures fall
within the {\it HST} field of view; as Figure~\ref{figcap13} shows,
Holes \#3 and \#8 have no detailed stellar population information
available.

\begin{figure}
\epsscale{1.0}
\plotone{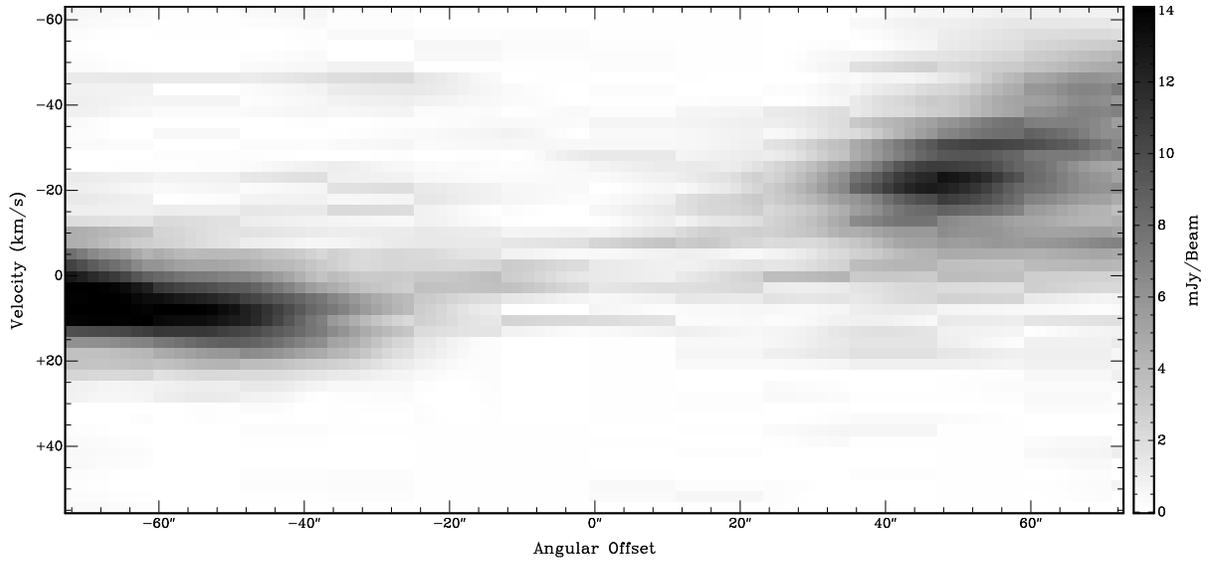}
\epsscale{1.0}
\caption{Major-axis PV slice through Hole \#6 (a Type~1 hole,
with little to no gas contained within).  No kinematic information is
available for a Type~1 hole.}
\label{figcap14}
\end{figure}

The most easily identifiable structure is the giant \HI\ hole
(identified as \#6 in Table~\ref{t4} and in Figures~\ref{figcap12} and
\ref{figcap13}).  This appears as an obvious evacuated cavity in
surface density maps and in the spatially resolved PV cuts (see
discussion in \S~\ref{S4.1.1}).  A PV cut through Hole \#6 is shown in
Figure~\ref{figcap14}; note the evacuated nature of the hole.  RV
analysis of this structure shows no expansion signature, and we thus
classify this structure as a Type~1 hole.  Note that the major axis of
the hole is aligned with the major axis used to create the PV slices
shown in Figures~\ref{figcap10} and \ref{figcap11}.  As
Figure~\ref{figcap13} shows, this hole occupies a significant fraction
of the entire galaxy, measuring 2.2\,$\times$\,1.1 kpc in diameter.
The stellar density within the hole is non-uniform, with areas of
higher density in the southeast and west, and sampling the transition
into the halo in the north.

We identify two other Type~1 holes in the neutral ISM of DDO\,165.
Hole \#4 appears as a slight under-density in the \HI\ column density
map (see Figure~\ref{figcap12}).  It is located well south of the
characteristic stellar ``cutoff'' in the southern region of the galaxy
(see Figure~\ref{figcap13}).  It is the smallest structure that we
detect with confidence and is only resolved in the robust-weighted
datacubes.  Hole \#1 is a highly elliptical feature (nearly 1 kpc in
length but only $\sim$400 pc wide) located in the northwest region of
the system.  The stellar density is somewhat lower than that within
Hole \#6.

\begin{figure}
\epsscale{1.0}
\plotone{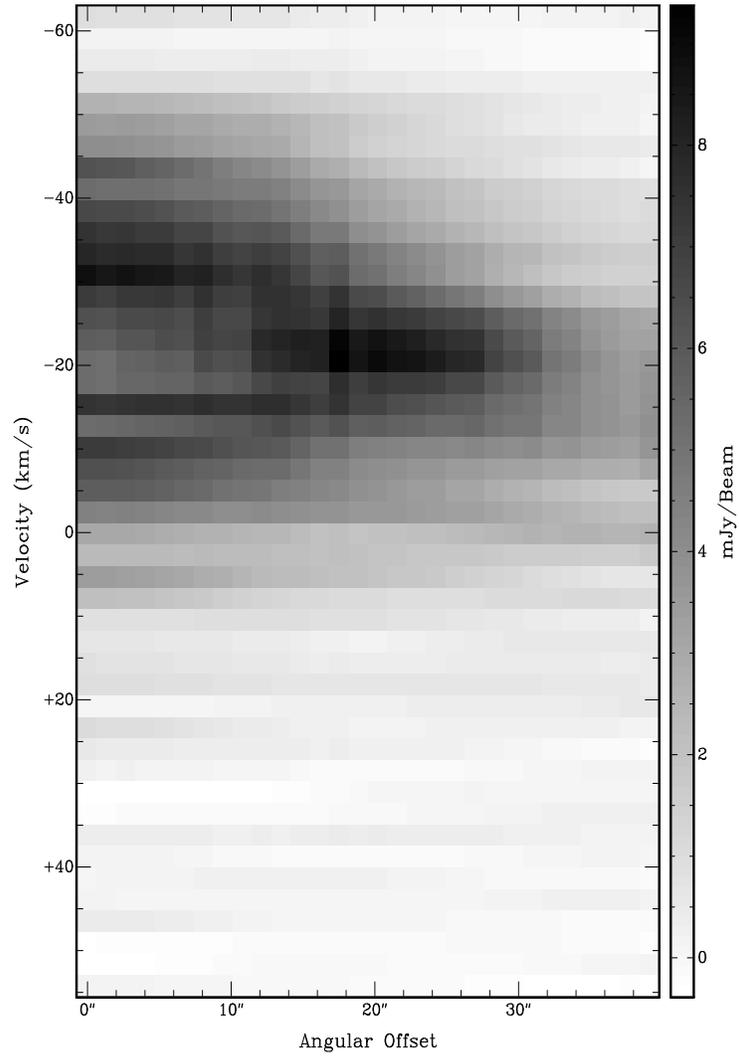}
\epsscale{1.0}
\caption{RV diagram of Hole \#5 (a Type~3 hole). The elliptical shell
is clearly visible; the radius of the structure is measured as the angular 
offset, and the expansion velocity is measured as half of the ellipse 
extent in the vertical direction.}
\label{figcap15}
\end{figure}

We identify five Type~3 holes in our analysis; Holes \#2 and \#3 are
apparent in both PV and RV space, while Holes \#5, \#7, and \#8 are
only found in the RV plane.  Holes \#2 and \#8 share similar
characteristics; both are located on the edge of a large \HI\ column
density maximum (but offset from the regions of high \halpha\
equivalent width).  Figure~\ref{figcap13} shows an apparent
over-density of stars near the center of Hole \#2; Hole \#8 falls
outside of the {\it HST} field of view.  Hole \#3 is located in the
outer disk in a region of comparatively low \HI\ surface density; it
is the largest expanding structure that we identify in this analysis.
Holes \#5 and \#7 are located in similar regions, immediately to the
south of the main stellar disk, just overlapping the transition region
to high surface densities (see Figure~\ref{figcap13}).  Both
structures are coincident with some of the high velocity \HI\ features
identified in \S~\ref{S4.1.2}.  As an example of the clarity of the
expansion signal in the RV plane, we show in Figure~\ref{figcap15} an
RV diagram of Hole \#5; note the closed half-ellipse structure.  The
diameter of the structure is estimated to be 17\arcsec, and the
minimum velocity is estimated to be $\sim$7 \kms.  We estimate the
errors on these measurements to be at the $\sim$30\% level.

The five structures with kinematic information span a range of
physical sizes (from $\sim$400 pc to $\sim$900 pc) and measured
expansion velocities (7 \kms\ to 11 \kms).  We stress that coherent
neutral gas structures can elude our detection methods in three ways.
First, expanding structures may have expansion velocities below our
2-channel velocity resolution limit (5.12 \kms).  Second, the
structures can be smaller than our reliable physical resolution limit
($\sim$165 pc from the 7.5\arcsec\ datacube).  Finally, and perhaps
most importantly, the expansion signature can be taking place out of
our line of sight.  We speculate that this effect may in part explain
the lack of Type~2 holes recovered by our analysis.

A region of significant interest is the northern \HI\ component;
recall the kinematic discontinuity between this region and the
southern \HI\ component (see discussion in \S~\ref{S4.1.1}).  The
northern \HI\ component harbors a visually identified stellar
association and ongoing SF (see Figures~\ref{figcap6} and
\ref{figcap7}).  We searched for signatures of expansion caused by
localized feedback from this SF, but none were found.  The separation
of the northern and southern \HI\ components, both physically and in
velocity space, is apparently not the result of feedback from recent
SF in the northern region alone.  We consider this issue in more
detail in Paper~II.

Note that we do not detect holes or shells directly associated with
either of the \halpha-bright SF regions in the main disk (compare
Figures~\ref{figcap13} and \ref{figcap5}).  While these SF regions are
coincident with high velocity gas features (see discussion in
\S~\ref{S4.1.2}), the recent ($\lsim$10 Myr) feedback from massive
stars has not produced coherent \HI\ structures large enough to be
prominent at our resolution limit.  We postulate that in these
regions, the mechanical energy from SF has increased the local
velocity dispersion of the gas, producing incoherent gas motions that
are extended in velocity space but that are not identifiable as
coherent \HI\ holes or shells.

\section{A Simplistic Estimate of \HI\ Structure Energetics}
\label{S5}

The coherent \HI\ structures that show evidence for expansion offer an
opportunity to derive the amount of energy needed to create the
structure.  The basic approach is to adopt the observed expansion
velocity as the constant, past expansion velocity. One then uses the
physical radius of the structure to infer a kinematic age, and an
estimate of the mass of gas within it (i.e., assuming a volume density
of neutral hydrogen) to derive the amount of energy needed to evacuate
the region.  This energy may originate from SNe and stellar winds, and
may be tied fundamentally to the recent SF patterns in a given region
of a galaxy.  Many past investigations \citep[e.g.,][]{walter99} have
adopted this formalism to infer the energetics of holes and shells in
external galaxies.

Most applications of this formalism use the single-blast model of 
\citet{chevalier74}.  Here, the energy of an expanding \HI\ shell is 
given by

\begin{equation}
{\rm E_{hole} = 5.3\times10^{43}}\, {\rm n_0^{1.12}}\, {\rm (\frac{d}{2})^{3.12}}\, {\rm v_{exp}^{1.4}} \ {\rm ergs},
\label{eq1}
\end{equation}

\noindent where E$_{\rm hole}$ is the energy needed to create the
expanding structure in ergs, n$_{\rm 0}$ is the \HI\ volume density
(for our calculations we assume n$_{\rm 0}$=0.1 cm$^{-3}$, an
appropriate average value for the outer disks of dwarf galaxies and
for DDO\,165; see detailed discussion in Paper~II), d is the diameter
of the shell in parsecs, and v$_{\rm exp}$ is the observed expansion
velocity in \kms.  This simplistic model has well-known difficulties;
some of the most pronounced include the unknown conversion efficiency
of SNe energy into gas momentum (assumed to be 100\% in a direct
application of Equation~\ref{eq1}), the explicit values of hydrogen
volume density as a function of position, the shape and depth of the
gravitational potential well prior to the creation of the hole, among
many others.  We stress that the above estimate is uncertain at the
50\% level or more; however, it is used frequently in the literature
and we thus include its application here for ease of comparison with
previous works.  Following {Weisz \etal\ (2009b)}\nocite{weisz09b} and
references therein, we revise some of the assumptions above to attempt
a more accurate calculation of the \HI\ hole energetics in Paper~II.

For the Type~3 holes identified in DDO\,165, we list in Table~\ref{t4}
their kinematic ages in Myr and the energetic requirements derived
using Equation~\ref{eq1}.  These energies can be compared to the
canonical energy output of a Type~II SN explosion, $\sim$10$^{51}$ erg
\citep{burrows00}.  One can thus infer the number of massive stars
required to create a given hole.  We note that the energies derived in
Table~4 are uncertain due to the assumption of a constant expansion
velocity.  The kinematic ages range from $\sim$20--50 Myr, while the
energy requirements are in the $\sim$10$^{51}$--10$^{52}$ erg range.
These values are in good agreement with the creation energies derived
in previous studies of \HI\ holes in dwarf galaxies (see, e.g.,
{Walter \& Brinks 1999}\nocite{walter99}, {Weisz
  \etal\ 2009b}\nocite{weisz09b}, and references therein).  Taken at
face value, these modest energy requirements suggest that the creation
of coherent \HI\ holes and shells can be accomplished by modest
numbers of massive stars.  However, the small number of detected
holes, and the extended recent SF throughout DDO\,165 (see {Weisz
  \etal\ 2008}\nocite{weisz08}), cloud this simplistic interpretation
and provide the motivation for the comparison of \HI\ and stellar
energetics that is the focus of Paper~II.

The lack of coherently expanding neutral gas structures associated
with the \halpha-luminous regions in DDO\,165 can be explained by
three possible scenarios.  First, the timescale for the creation of
coherent \HI\ structures from small numbers of massive stars could be
longer than the characteristic timescale for \halpha\ emission (10
Myr).  Second, the ongoing SF in DDO\,165, concentrated in the
\halpha-luminous regions, could be too weak to create such structures
at all; the efficiency of individual SNe at accelerating natal neutral
gas could be very low.  Third, if that efficiency is low, then the
creation of such structures might require SF that is extended
temporally in order to inject the requisite amounts of energy.
Evidence is mounting that these long burst episodes are important in
the ISM of dwarf galaxies. Elevated SFRs have been measured to span
hundreds of Myr, much longer than the instantaneous SF traced by
ionizing photons ({McQuinn \etal\ 2009}\nocite{mcquinn09},
{2010a}\nocite{mcquinn10a}, {2010b}\nocite{mcquinn10b}).  These
temporally extended SF episodes have been shown to be energetically
capable of creating the \HI\ holes observed in nearby galaxies ({Weisz
  \etal\ 2009a}\nocite{weisz09a}, {2009b}\nocite{weisz09b}).  Taken
together, these lines of evidence suggest that it is not surprising
that the holes and shells we observe in DDO\,165 are in fact not
associated with the currently most massive stars.

While the expanding holes in DDO\,165 provide important kinematic
information, the Type~1 holes are equally intriguing.  We again draw
attention to Hole \#6, the giant structure that spans a
$\sim$2.2\,$\times$\,1.1 kpc region offset from the center of the
stellar body.  This structure has been nearly completely evacuated of
neutral gas (N$_{\rm HI}$ $<$10$^{20}$ cm$^{-2}$ at
10\arcsec\ resolution); given the number of young stars located within
the hole (see Paper~II for details), it is logical to conclude that
whatever mechanism conspired to evacuate it did so during or after the
formation of this interior stellar population.  We note that the blue
stellar density distribution presented in \citet{weisz08} contains a
significant number of stars in both the southern \HI\ component and in
the area now covered by the giant \HI\ hole.  While the estimates of
the requisite energetics are difficult in the absence of kinematic
information, we nonetheless explore the connection between recent SF
and this giant \HI\ structure in more detail in Paper~II.

\section{Conclusions}
\label{S6}

We have presented new VLA multi-configuration \HI\ 21~cm spectral line
observations of the intriguing post-starburst dwarf irregular galaxy
DDO 165.  The neutral gas morphology and dynamics are quite complex
for an apparently isolated low-mass system.  The two major \HI\
components appear to be kinematically distinct.  The southern \HI\
component contains the bulk of the neutral gas mass ($\sim$85\% or
9.4\,$\times$\,10$^7$ \msun) and is spatially coincident with the
regions of highest stellar density and with the characteristic
``cutoff'' of the stellar population in the southern disk.  Two column
density peaks in this southern \HI\ component are spatially coincident
with the highest surface brightness \halpha\ emission.  The northern
component also harbors dense gas and ongoing SF.

The \HI\ morphology is dominated by a giant evacuated hole that spans
$\sim$2.2\,$\times$\,1.1 kpc.  This feature is offset from the region
of highest stellar density, but still contains a significant portion
of the entire stellar population of the DDO\,165 system.  In addition
to this visually obvious \HI\ hole, we find seven other hole or shell
features via PV or RV analysis.  Five of these features have kinematic
information (i.e., measured expansion velocities), while the other
three (including the aforementioned giant \HI\ hole) are currently
static in the \HI\ spectral line.  A coarse estimate of the requisite
energetics is performed using the oft-cited single-blast, kinematic
age model \citep{chevalier74}; Paper~II presents a more sophisticated
derivation of the energetic requirements of the \HI\ holes based on a
mass model derived from infrared imaging.

A spatially resolved PV analysis shows multiple high velocity gas
features along the southern \HI\ component, spatially coincident with
the truncated stellar population.  These localized features span an
average velocity range of $\sim$40 \kms, with some features reaching
$\sim$60 \kms.  These features are interpreted as the localized
effects of stellar feedback that have created incoherent motion of the
surrounding \HI\ gas; this hypothesis is explored in detail in
Paper~II.  This interpretation is strengthened by the results of the
spatially resolved SF history analyses of DDO\,165 presented in
\citet{weisz08} and in McQuinn \etal\ ({2010a}\nocite{mcquinn10a},
{2010b}\nocite{mcquinn10b}); the past SF was extended over hundreds of
Myr, and we postulate that feedback from this SF has produced the
highly disturbed \HI\ dynamics of the system.

Using the kinematic information in our \HI\ datacubes, we can reject
certain models for the origin of the two kinematically distinct \HI\
components.  Specifically, scenarios involving turbulence,
counter-rotating or superposed disks, and coherent global rotation
each fail to reproduce the kinematic signatures in our data.  However,
we are unable to differentiate between an infall or outflow origin
using the \HI\ data alone; both mechanisms are considered further in
Paper~II.

DDO\,165 is at an important phase of its evolution.  The recent
temporally and spatially extended SF event has had a significant
effect on the surrounding ISM.  It thus provides an opportunity to
improve our understanding of galaxy evolution.  To further explore the
nature of SF in DDO\,165, and the capabilities of that SF to shape the
neutral ISM (e.g., creating high velocity gas features, holes and
shells of various sizes and energetic requirements, and potentially
the blowout of significant fractions of the total gas mass), we
present a detailed comparison of the \HI\ kinematics and the spatially
resolved SF history in Paper~II.

\acknowledgements
 
Partial support for this work was provided by NASA through grant
GO-10605 from the Space Telescope Science Institute, which is operated
by AURA, INC., under NASA contract NAS5-26555.  J.M.C. and
H.P.M. thank Macalester College for research support.  D.R.W. and
S.R.W. are grateful for support from Penrose Fellowships.  This
investigation has made use of the NASA/IPAC Extragalactic Database
(NED) which is operated by the Jet Propulsion Laboratory, California
Institute of Technology, under contract with the National Aeronautics
and Space Administration, and NASA's Astrophysics Data System.

\clearpage
\bibliographystyle{apj}                                                 


\clearpage
\begin{deluxetable}{lc}  
\tablecaption{Basic Parameters of DDO\,165} 
\tablewidth{0pt}  
\tablehead{ 
\colhead{Property}    
&\colhead{Values}}    
\startdata      
\vspace{0.0 cm} 
Alternate name  &UGC 8201 \\    
Right Ascension (J2000)   &13$^{\rm h}$ 06$^{\rm m}$ 24.$^{\rm s}$8\\        
Declination (J2000)     &+67\arcdeg 42\arcmin 25\arcsec\\    
Adopted distance\tablenotemark{a} (Mpc) &4.47 \pom\ 0.20\\
A$_{\rm I}$\tablenotemark{b} &0.047 \pom\ 0.010\\
S$_{\rm HI}$ (Jy km/s)\tablenotemark{c} &23.2 \pom\ 2.0\\
\HI\ mass (10$^8$ \msun)\tablenotemark{c} &1.1 \pom\ 0.3 \\
\enddata     
\label{t1}
\tablenotetext{a}{\citet{karachentsev02}}
\tablenotetext{b}{\citet{schlegel98}}
\tablenotetext{c}{This study}
\end{deluxetable}   
     
\clearpage                                     
\begin{deluxetable}{lccc}
\tablecaption{Setup of the VLA during Observations} 
\tablewidth{0pt}  
\tablehead{ 
\colhead{Parameter}    
&\colhead{B}
&\colhead{C}
&\colhead{D}}
\startdata      
\vspace{0.0 cm}      
Baselines &1-53 k${\lambda}$ &0.2-16.1 k${\lambda}$ &0.2-4.5 k${\lambda}$\\
Date of Observation &Nov 9, 2007 &Nov 19, 2006 &Apr 11, 2007\\
Total time on source (min) &390 &172 &85\\
Total Bandwidth (MHz) &1.56 &1.56 &1.56\\
Number of Channels &128 &128 &128\\
Velocity Resolution (\kms) &2.58 &2.58 &2.58\\
\enddata     
\label{t2}
\end{deluxetable}  

\clearpage
\begin{deluxetable}{lccc}
\tablecaption{\halpha\ regions in DDO\,165} 
\tablewidth{0pt}  
\tablehead{ 
\colhead{Number}    &\colhead{$\alpha$ (J2000)\tablenotemark{a}} &\colhead{$\delta$ (J2000)\tablenotemark{a}} &\colhead{Flux (erg\,s$^{-1}$\,cm$^{-2}$)}}
\startdata      
\vspace{0.0 cm}      
1  &13:06:11.5 &67:41:48.4 &2.60E-15\\
2  &13:06:17.5 &67:42:09.0 &3.69E-14\\
3  &13:06:19.8 &67:42:02.7 &7.25E-15\\
4  &13:06:20.8 &67:41:43.7 &4.35E-15\\
5  &13:06:22.5 &67:43:22.7 &2.45E-15\\
6  &13:06:23.3 &67:41:59.0 &7.05E-15\\
7  &13:06:24.5 &67:43:40.1 &1.23E-14\\
8  &13:06:27.2 &67:43:20.4 &2.15E-15\\
9  &13:06:38.0 &67:42:12.0 &2.58E-15\\
10 &13:06:41.8 &67:42:11.6 &7.37E-15\\
\enddata     
\label{t3}
\tablenotetext{a}{The Right Ascension and declination at the center of
the \halpha\ regions. The error of these values is \pom\ 1\arcsec. Units of Right
Ascension are hours, minutes, and seconds, and units of declination
are degrees, arcminutes, and arcseconds.}  
\end{deluxetable}  

\clearpage
\begin{deluxetable}{lcccccccc}
\tabletypesize{\scriptsize}
\tablecaption{Catalog of \HI\ Holes in DDO\,165} 
\tablewidth{0pt}  
\tablehead{ 
\colhead{Number}    &\colhead{$\alpha$} &\colhead{$\delta$} &\colhead{Size} &\colhead{$V_{\rm exp}$} &\colhead{Age t} &\colhead{Energy E$_{\rm hole}$} &\colhead{Type\tablenotemark{e}} &\colhead{Identification\tablenotemark{f}}\\
\colhead{}    &\colhead{(J2000)\tablenotemark{a}} &\colhead{(J2000)\tablenotemark{a}} &\colhead{(pc)\tablenotemark{b}} &\colhead{(km\,s$^{-1}$)\tablenotemark{c}} &(Myr)\tablenotemark{d} &\colhead{(10$^{51}$ ergs)} &\colhead{} &\colhead{}
}
\startdata      
\vspace{0.0 cm}      
1 &13:06:17.1 &67:43:04.4  &975 $\times$ 325, 40\degr &- &- &- &1 &PV\\
2 &13:06:22.1 &67:41:46.9 &520 &11 &23.0 &3.95 &3 &PV, RV\\
3 &13:06:23.3 &67:40:43.4 &867 &10 &42.2 &170.5 &3 &PV, RV\\
4 &13:06:27.7 &67:41:04.5  &390 &- &- &- &1 &PV\\
5 &13:06:30.1 &67:41:26.5 &737 &7 &51.2 &6.23 &3 &RV\\
6 &13:06:30.6 &67:42:31.0 &2210 $\times$ 1084, 130\degr &- &- &- &1 &PV\\
7 &13:06:38.6 &67:41:33.7 &455 &9 &24.6 &1.97 &3 &RV\\
8 &13:06:45.2 &67:42:02.1  &433 &7 &30.1 &11.9 &3 &RV\\
\enddata     
\label{t4}
\tablenotetext{a}{The Right Ascension and declination of the center of
the holes. The error of these values is \pom\ 3\arcsec. Units of Right
Ascension are hours, minutes, and seconds, and units of declination
are degrees, arcminutes, and arcseconds.}  
\tablenotetext{b}{The diameter of the hole in pc. If the hole is
elliptical then the major and minor axis diameters, and the major axis 
position angle (east of north), are also listed.  The hole diameters 
listed have an error of \pom\ 100 pc}
\tablenotetext{c}{Expansion velocity of the hole, when it can be
measured, with an error of 2 \kms.}
\tablenotetext{d}{The calculated ages for the holes are upper limits
as we assumed the expansion velocity to be constant for the holes
creation.}
\tablenotetext{e}{Hole type as discussed in \S4.1}
\tablenotetext{f}{The holes were either identified by a position
velocity (PV) slice, a radius velocity (RV) slice, or a combination of
the two.}
\end{deluxetable}  

\clearpage

\clearpage

\clearpage

\clearpage

\clearpage

\clearpage

\clearpage

\clearpage

\clearpage

\clearpage

\clearpage

\clearpage

\clearpage

\clearpage

\clearpage

\end{document}